 \documentclass[final,3p,times]{elsarticle}

\usepackage{amssymb}

\usepackage{amsmath}
\usepackage{caption}

\newcounter{bla}

\usepackage{array}
\usepackage{booktabs}
\usepackage{xcolor}

\newcommand{\gv}[1]{\ensuremath{\mbox{\boldmath$ #1 $}}} 
\newcommand{\guv}[1]{\ensuremath{\mbox{\boldmath$ \hat{#1} $}}} 

\begin{document}

\begin{frontmatter}



\title{Drift reduced Landau fluid model for magnetized plasma turbulence simulations in BOUT++ framework}


\author[a]{Ben Zhu\corref{author}}
\author[b]{Haruki Seto}
\author[a]{Xue-qiao Xu}
\author[b]{Masatoshi Yagi}

\cortext[author] {Corresponding author.\\\textit{E-mail address:} zhu12@llnl.gov}
\address[a]{Lawrence Livermore National Laboratory, Livermore, California 94550, USA}
\address[b]{National Institutes for Quantum and Radiological Science and Technology, Rokkasho, Aomori 039-3212, Japan}

\begin{abstract}
Recently the drift-reduced Landau fluid six-field turbulence model within the BOUT++ framework~\cite{dudson2009bout++} has been upgraded. In particular, this new model employs a new normalization, adds a volumetric flux-driven source option, the Landau fluid closure for parallel heat flux and a Laplacian inversion solver which is able to capture $n=0$ axisymmetric mode evolution in realistic tokamak configurations. These improvements substantially extended model's capability to study a wider range of tokamak edge phenomena, and are essential to build a fully self-consistent edge turbulence model capable of both transient (e.g., ELM, disruption) and transport time-scale simulations.

\end{abstract}


\end{frontmatter}

\section{Introduction}

Magnetized plasma is of great interest to scientists and engineers for its applications in astrophysics, space physics, material processing and fusion energy research. In this paper, we describe a fluid model that aims to study the instability, turbulence, and transport for magnetically confined plasmas. 
In particular, this model is designed to investigate the rich physics on the periphery of a tokamak (a toroidal device that confines high temperature plasmas with a swirling magnetic field) plasma, commonly termed as the edge plasma. The edge plasma is imperative to the success of self-sustained fusion operations as it largely controls the particle and energy exhaust, breeding and refueling, and overall fusion performance. It therefore becomes one of the most active research areas in fusion plasma and engineering nowadays.

The tokamak edge region is a complicated physics system. It normally has a radial extension of a few centimeters and consists of two distinct magnetic topologies -- the closed flux region where magnetic field-lines keep circling around major axis of the torus, and the scrape-off-layer (SOL) as well as the private flux region where magnetic field-lines intersect with vessel wall and/or divertor targets. Within such a short radial distance, plasma density and temperature could drop one or two orders of magnitude, transforming from a hot, dense, nearly collisionless core plasma to a relatively cold, thin, collisional SOL plasma. The resulting strong radial inhomogeneity of plasma pressure further modifies local plasma current density profile and hence the local magnetic shear, and could excite various plasma instabilities ranging from the large size magnetohydrodynamic (MHD) instabilities to the less harmful micro-instabilities at ion and electron gyroraduis scales. 

These instabilities, usually cascade into turbulence, enhance cross-field transport, alter the equilibrium profiles and degrade fusion performance. 
The typical edge turbulence fluctuation level may vary from $O(10^{-2})$ to order of unity~\cite{francisquez2017global} and in most cases, is largely impacted by local plasma conditions.
However, global simulations suggest that fully developed edge turbulence can also be non-local, i.e., turbulence can have long-range influence on either up-gradient~\cite{ma2015impact} or down-gradient~\cite{chen2018progress} or both regions via turbulence spreading~\cite{hahm2004turbulence}, depending on the nature of the underlying instability.
In the meantime, strongly sheared flows occurred the edge region (e.g., the $E\times B$ flow in the poloidal direction and the intrinsic and/or driven rotation in the toroidal direction) are found to be crucial in terms of governing the overall edge plasma performance. Although the interaction between the sheared flows and turbulence saturation has been extensively investigated in theoretical, experimental and numerical studies, the spontaneous generation and saturation mechanisms are still not fully understood.

Last but not least, at the far-SOL and near the wall where plasma temperature is relatively cool, impurity, neutral and molecular physics become important as the atomic and molecular processes (e.g., radiation, recombination, ionization, charge-exchange, dissociation, etc) could dramatically change the local plasma dynamics and further affect upstream plasma.
Therefore, tokamak edge is an inherent multi-scale, multi-physics, highly nonlinear system that requires sophisticated numerical modelling.

In recent years, a large effort has been devoted to tokamak edge modelling and significant progress has been made. Several fluid (e.g., GBS~\cite{ricci2012simulation}, TOKAM3X~\cite{tamain2016tokam3x}, GDB~\cite{zhu2018gdb}, GRILLIX~\cite{stegmeir2018grillix}), gyro-fluid (e.g., GEMR~\cite{scott2006edge} BOUT++-GLF~\cite{ma2016global}) and gyrokinetic (e.g., XGC~\cite{chang2004numerical}, Gkeyll~\cite{hakim2020continuum}, COGENT~\cite{dorf2020progress}) codes have been developed to investigate different aspects of the plentiful edge related problems.
However, the extremely demanding computational requirement of these high-fidelity multi-scale gyrokinetic models limits their applications in the routine global edge simulations.
Concurrently, fluid models, though only contain the lowest three moments of plasma distribution functions, still capture the fundamental physics and are adequate to quantitatively describe the behavior of edge plasmas in many cases. For this reason, they are and will remain an irreplaceable component of the edge study~\cite{francisquez2020fluid}.

The original Boundary Turbulence (BOUT) fluid code was first developed in the late 1990s~\cite{xu1998scrape,xu2008boundary} to explore turbulence dynamics at the tokamak boundary region (i.e., the edge). It was continuously refined over the years~\cite{umansky2009status} and eventually extended to BOUT++~\cite{dudson2009bout++} -- an open source C++ framework that provides a user friendly interface for physicists to code partial differential equations (PDEs) without worrying about numerical implementation. This successful strategy yields numerous numerical models targeted for specific research topics within the BOUT++ framework to explore a wide range of edge physics – from linear growth rate analysis to fully nonlinear, full-$f$ turbulence simulation; from quasi-state transport to intermittent events such as ELM crashing; from particle fueling, radio frequency (RF) heating to divertor exhaust solutions. Nevertheless, the six-field turbulence model in BOUT++ framework~\cite{xia2013six} is one of the main workhorses among all BOUT++ models. Recently, it has been substantially improved -- it now simulates a more complete physics model with a more practical normalization and implies a zonal field solver, a sophisticated Landau fluid parallel heat flux model and a flux-driven source option.

This paper is by no means to provide a complete manual of the drift-reduced turbulence model in BOUT++, but rather a self-contained documentation that reflects current status of the model with a focus on the physics aspect to facilitate future research. The rest of the paper is organized as follows. Section~\ref{sec:equations} introduces the equations and the normalization used in the BOUT++ turbulence model. Field-aligned coordinate system that minimizes computational costs and the associated operators, as well as the boundary treatment are outlined in Section~\ref{sec:coordinate}.
Section~\ref{sec:lf} presents the latest developed Landau fluid closure on parallel heat flux for weakly collisional tokamak edge plasma in divertor configuration. The Landau fluid closure prediction is compared and contrasted with those from the widely used classical Braginskii and flux-limited models.
The inversion of Laplacian operator, for both non-axisymmetric ($n\neq0$) and axisymmetric ($n=0$) components, is described in Section~\ref{sec:zonal}. Section~\ref{sec:source} illustrates the application of flux-driven source. Finally, Section~\ref{sec:sum} summaries the current status of the drift-reduced Landau fluid model in BOUT++ and discusses the on-going and planned efforts on future model development.


\section{Model equations}\label{sec:equations}
Like the other fluid turbulence models of its kind, BOUT++ turbulence model is also based on the seminal Braginskii description of collisional, magnetized plasma~\cite{braginskii1965transport}. In particular, BOUT++ implements the two-fluid drift-reduced Braginskii equations~\cite{simakov2003drift,xu2008boundary} which eliminate high frequency dynamics (e.g., gyration and compressional Alfv\'en wave) by assuming that $E\times B$ and diamagnetic drifts are the lowest order flows retained in system, and therefore is suitable to investigate low frequency ($\omega\ll\omega_{ci}$) turbulence and transport phenomena in the magnetized plasma. 
As will be discussed in Section~\ref{sec:lf} the collisionality constraint can be relaxed to some extent once the original Braginskii parallel heat flux model is replaced by the Landau fluid closure (and hence the name Landau fluid model). 

\subsection{Plasma equations}
BOUT++'s drift-reduced Landau fluid model describes the ion scale, low frequency turbulence in a relatively collisional, magnetized, quasi-neutral ($n_e=Zn_i$) plasma. Its equation set consists six independent primitive variables -- ion density $n_i$, electrostatic potential $\phi$, ion parallel velocity $V_{\parallel,i}$, perturbed (parallel) magnetic flux $A_\parallel$, electron and ion temperature $T_{e,i}$, according to
\begin{align}
\label{eq:density}
\frac{\partial}{\partial t}n_{i} &= -\left(\frac{1}{B}\boldsymbol{\hat{b}}\times\boldsymbol{\nabla}_{\perp}\phi+V_{\parallel i}\boldsymbol{\hat{b}}\right)\cdot\boldsymbol{\nabla} n_{i}
-\frac{2n_{i}}{B}\boldsymbol{\hat{b}}\times\boldsymbol{\kappa}\cdot\boldsymbol{\nabla}_{\perp}\phi+\frac{2}{ZeB}\boldsymbol{\hat{b}}\times\boldsymbol{\kappa}\cdot\boldsymbol{\nabla}_{\perp}P_{e}-n_{i}B\nabla_{\parallel}\left(\frac{V_{\parallel i}}{B}\right)+\frac{B}{Ze}\nabla_\parallel\left(\frac{J_\parallel}{B}\right)+S_n,\\
\frac{\partial}{\partial t}\varpi & = -\left(\frac{1}{B}\boldsymbol{\hat{b}}\times\boldsymbol{\nabla}_{\perp}\phi+V_{\parallel i}\boldsymbol{\hat{b}}\right)\cdot\boldsymbol{\nabla}\varpi + B^{2}\boldsymbol{\nabla}_\parallel \left(\frac{J_{\parallel}}{B}\right)+2\boldsymbol{\hat{b}}\times\boldsymbol{\kappa}\cdot\boldsymbol{\nabla}P -\frac{2}{3}\boldsymbol{\hat{b}}\times\boldsymbol{\kappa}\cdot\boldsymbol{\nabla}\pi_{ci}\nonumber \\
& -\frac{1}{2\Omega_{i}}\left[n_{i}Ze\boldsymbol{V_{D_i}}\cdot\boldsymbol{\nabla}\left(\nabla_{\perp}^{2}\phi\right)-m_{i}\Omega_{i}\boldsymbol{\hat{b}}\times\boldsymbol{\nabla}n_{i}\cdot\boldsymbol{\nabla}V_{E}^{2}\right] +\frac{1}{2\Omega_{i}}\left[\boldsymbol{V_E}\cdot\boldsymbol{\nabla}\left(\nabla_{\perp}^{2}P_{i}\right)-\nabla_{\perp}^{2}\left(\boldsymbol{V_E}\cdot\boldsymbol{\nabla}P_{i}\right)\right], \\
\frac{\partial}{\partial t}V_{\parallel i} &= -\left(\frac{1}{B}\boldsymbol{\hat{b}}\times\boldsymbol{\nabla}_{\perp}\phi+V_{\parallel i}\boldsymbol{\hat{b}}\right)\cdot\boldsymbol{\nabla}V_{\parallel i}-\frac{1}{m_{i}n_{i}}\nabla_\parallel P
-\frac{2}{3m_{i}n_{i}}B^{\frac{3}{2}}\nabla_{\parallel}\left(\frac{\pi_{ci}}{B^{\frac{3}{2}}}\right)-\boldsymbol{V_{D_i}}\cdot\boldsymbol{\nabla}V_{\parallel i}-\frac{V_{\parallel i}S_n}{n_i},\\
\frac{\partial}{\partial t}A_\parallel^* &= -\nabla_{\parallel}\phi+\frac{\eta_{\parallel}}{\mu_{0}}\nabla_{\perp}^{2}A_\parallel+\frac{1}{en_{e}}\nabla_{\parallel}P_{e}+\frac{0.71k_{B}}{e}\nabla_{\parallel}T_{e},\\
\frac{\partial}{\partial t}T_{i} & = -\left(\frac{1}{B}\boldsymbol{\hat{b}}\times\boldsymbol{\nabla}_{\perp}\phi+V_{\parallel i}\boldsymbol{\hat{b}}\right)\cdot\boldsymbol{\nabla}T_{i} +\frac{2}{3n_{i}k_{B}}\nabla_{\parallel}q_{\parallel i} +\frac{2m_{e}}{m_{i}}\frac{Z}{\tau_{e}}\left(T_{e}-T_{i}\right) \nonumber \\
& -\frac{2}{3}T_{i}\left[\left(\frac{2}{B}\boldsymbol{\hat{b}}\times\boldsymbol{\kappa}\right)\cdot\left(\boldsymbol{\nabla}\phi+\frac{1}{Zen_{i}}\boldsymbol{\nabla}P_{i}+\frac{5}{2}\frac{k_{B}}{Ze}\boldsymbol{\nabla}T_{i}\right)+B\nabla_{\parallel}\left(\frac{V_{\parallel i}}{B}\right)\right] \nonumber \\
 & -\frac{2\pi_{ci}}{9k_{B}n_{i}}\left[\frac{2}{\sqrt{B}}\nabla_{\parallel}\left(\sqrt{B}V_{\parallel i}\right)-\frac{k_{B}}{Zen_{i}B}\boldsymbol{\hat{b}}\cdot\boldsymbol{\nabla}n_{i}\times\boldsymbol{\nabla}T_{i}\right]-\frac{4}{3\Omega_{i}}T_{i}V_{\parallel i}\boldsymbol{\hat{b}}\times\boldsymbol{\kappa}\cdot\boldsymbol{\nabla}V_{\parallel i}+\frac{2 S^E_i}{3n_i}-\frac{T_i S_n}{n_i}, \\
\label{eq:ti}
\frac{\partial}{\partial t}T_{e} & = -\left(\frac{1}{B}\boldsymbol{\hat{b}}\times\boldsymbol{\nabla}_{\perp}\phi+V_{\parallel e}\boldsymbol{\hat{b}}\right)\cdot\boldsymbol{\nabla}T_{e} + \frac{2}{3n_{e}k_{B}}\nabla_{\parallel}q_{\parallel e} +0.71\frac{2T_{e}}{3en_{e}}B\nabla_{\parallel}\left(\frac{J_{\parallel}}{B}\right) -\frac{2m_{e}}{m_{i}}\frac{1}{\tau_{e}}\left(T_{e}-T_{i}\right) \nonumber \\
& -\frac{2}{3}T_{e}\left[\left(\frac{2}{B}\boldsymbol{\hat{b}}\times\boldsymbol{\kappa}\right)\cdot\left(\boldsymbol{\nabla}\phi-\frac{1}{en_{e}}\boldsymbol{\nabla}P_{e}-\frac{5}{2}\frac{k_{B}}{e}\boldsymbol{\nabla}T_{e}\right)+B\nabla_{\parallel}\left(\frac{V_{\parallel e}}{B}\right)\right] +\frac{2}{3n_{e}k_{B}}\eta_{\parallel}J_{\parallel}^{2}+\frac{2 S^E_e}{3n_e}-\frac{T_e S_n}{n_e}.
\end{align}

Here vorticity $\varpi$ is defined as
\begin{equation}\label{eq:vort}
    \varpi= \frac{n_i m_i}{B}\left(\nabla_\perp^2\phi +\frac{1}{n_i}\boldsymbol{\nabla}_\perp\phi\cdot\boldsymbol{\nabla_\perp} n_i+\frac{1}{Zen_i}\nabla_\perp^2P_i\right),
\end{equation}
magnetic curvature
\begin{equation}
    \boldsymbol{\kappa}=\boldsymbol{\hat{b}}\cdot\boldsymbol{\nabla}\boldsymbol{\hat{b}}\simeq
    \boldsymbol{\hat{b}}_0\cdot\boldsymbol{\nabla}\boldsymbol{\hat{b}}_0 \quad\text{with}~~ \boldsymbol{\hat{b}}_0=\boldsymbol{B}_0/B_0,
\end{equation}
modified magnetic flux
\begin{equation}
    A_\parallel^*=\left(1-d_e^2\nabla_\perp^2\right)A_\parallel \quad\text{with}~~ d_e^2=c^2/\omega_{pe}^2,
\end{equation}
total parallel current density $J_\parallel=J_{\parallel 0}+j_\parallel$ with perturbed current density $j_\parallel=-\nabla_\perp^2A_\parallel/\mu_0$,
and ion parallel viscosity
\begin{equation}
    \pi_{ci}=\eta_i^0\left[ \boldsymbol{\kappa}\cdot (\boldsymbol{V_E}+\boldsymbol{V_{D_i}})- \frac{2}{\sqrt{B_0}}\partial_\parallel(\sqrt{B_0}V_{\parallel i})\right].
\end{equation}
The leading ion viscosity coefficient $\eta_i^0=0.96nk_BT_i\tau_{ii}$, the parallel electrical conductivity is $\eta_\parallel=F(Z)m_e/(ne^2\tau_e)$ with $F(Z)=(1+1.198Z+0.222Z^2)/(1+2.996Z+0.753Z^2)$ for arbitrary $Z$, the $E\times B$ and ion diamagnetic drifts are 
\begin{equation}
    \boldsymbol{V_E}=\boldsymbol{\hat{b}}\times \boldsymbol{\nabla}_\perp \phi/B\simeq\boldsymbol{\hat{b}}_0\times \boldsymbol{\nabla}_\perp \phi/B_0, \quad \boldsymbol{V_{D_i}}=\boldsymbol{\hat{b}}\times \boldsymbol{\nabla}_\perp P_i/(Zen_iB)\simeq\boldsymbol{\hat{b}}_0\times \boldsymbol{\nabla}_\perp P_i/(Zen_iB_0).
\end{equation}

Here we assume that perturbed magnetic field $\tilde{\boldsymbol{B}}\simeq \nabla A_\parallel \times \guv{b}_0$ is orders of magnitude smaller than the equilibrium magnetic field $\boldsymbol{B}_0$, a condition normally true in magnetically confined plasma, such that $|\tilde{\boldsymbol{B}}|\ll|\boldsymbol{B}_0|$, or $\boldsymbol{B}\approx \boldsymbol{B}_0$.

Under the drift-reduce ordering assumption, ion polarization drift is retained to leading order in order to ensure energy conservation even though its amplitude is small comparing to $E\times B$ and ion diamagnetic drifts. However, electron polarization drift along with other terms smaller by the order of $m_e/m_i$ are discarded during the derivation.
The electron and ion parallel heat flux $q_{\parallel e,i}$ will be discussed in detail in Section~\ref{sec:lf} as BOUT++ has implemented three different parallel heat flux models. The electron and ion perpendicular heat flux are often neglected in edge turbulence models since they are on the order of $(\omega_{cj}\tau_j)^{-2}\ll 1$ smaller than the parallel heat flux in the collisional magnetized plasmas. Nevertheless, including cross field neoclassical thermal diffusion effects~\cite{wesson2011tokamaks} in toroidal plasma simulations is available in this model.
The BOUT++ Landau fluid turbulence model has a build-in option to either consistently evolve or maintain the axisymmetric plasma profiles. The formal option is called non-scale separated, or full-field model since the background plasma profiles tend to relax once the turbulence is developed and starts to continually transport particles and energy radially outward. The external particle and energy sources terms $S_n,S_i^E,S_e^E$ are hence applied when the model is used for transport time scale global full-field turbulence simulation. The latter option, on the other hand, results in the scale separated, or gradient-driven turbulence model as the gradients of the background plasma profiles which excite instabilities and turbulence are fixed over the time. These two types of model are also often refereed as full-$f$ and $\delta f$ fluid models in a loose manner as analogous to kinetic simulations.

It is therefore reasonable and necessary to refuel particles $S_n$ and energy $S_i^E,S_e^E$ in order to prevent the collapse of plasma profiles in transport time scale simulations. As a result, the external particle and energy flux could largely alter the local plasma profiles and further influence the plasma dynamics both in turbulence and transport time scales, so such a model is also termed as the flux-driven turbulence model.

The equation set (\ref{eq:density})-(\ref{eq:ti}) is in fact a super-set of various transport and turbulence models built upon BOUT++ framework. For instance, if one sets $\partial A_\parallel^*/\partial t=0$, it becomes a five-field electrostatic turbulence model~\cite{xu1998scrape,xia2013five}; if one combines density and temperature equations into total plasma pressure equation, it becomes three-field model~\cite{xi2013impact}; if one makes electrostatic and axisymmetric assumptions (i.e.,$\partial/\partial \xi=0$ where $\xi$ is the toroidal direction) and neglected subdominant terms, it becomes 2D transport model~\cite{wang20142d}. 

\subsection{Normalization}
In the six-field model, basic normalization coefficients, e.g., the characteristic length $\bar{L}$, magnetic field strength $\bar{B}$, density $\bar{n}$, electron and ion temperature $\bar{T}_{e,i}$, are prescribed based on initial profiles. For most of tokamak simulations, $\bar{L},\bar{B}$ and $\bar{n}$ are the largest radius, total magnetic field strength and ion density respectively; $\bar{T}_{e,i}$ are typically set to be $1000eV$. With defined reference density and temperature, the individual pressure then can be normalized as $\bar{P}_j=\bar{n}k_B\bar{T}_{j}$, and the total pressure $\bar{P}=\bar{n}k_B\bar{T}_e$, $\hat{P}=\hat{P}_e+\tau\hat{P}_i$ with $\tau=\bar{T}_i/\bar{T}_e$. Similarly, $\boldsymbol{\hat{\kappa}}=\bar{L}\boldsymbol{\kappa}, \boldsymbol{\hat{\nabla}}=\bar{L}\boldsymbol{\nabla}$, and the characteristic velocity and time are
\begin{equation}
    \bar{V}_A=\bar{B}/\sqrt{\mu_0m_i\bar{n}},\quad \bar{t}=\bar{L}/\bar{V}_A.
\end{equation}

For the electromagnetic quantities, BOUT++ chooses a normalization that minimizes the number of normalization coefficients as
\begin{equation}
    \bar{\phi}=\bar{L}^2\bar{B}/\bar{t},\quad \bar{E}=\bar{B}\bar{V}_A,\quad \bar{\varpi}=m_i\bar{n}/\bar{t},\quad \bar{J}=\bar{B}/(\mu_0\bar{L}),\quad \bar{A}_\parallel=\bar{L}\bar{B}, \quad \bar{\eta}=\mu_0\bar{L}\bar{V}_A.
\end{equation}
Note that in this normalization, electric field, $E\times B$ and ion diamagnetic drifts are simply
\begin{equation}\label{eq:v_raw_norm}
    \boldsymbol{\hat{E}}=-\boldsymbol{\hat{\nabla}}\hat{\phi},\quad
    \boldsymbol{\hat{V}_E}=\frac{\boldsymbol{\hat{b}}\times\boldsymbol{\hat{\nabla}}\hat{\phi}}{\hat{B}},\quad
    \boldsymbol{\hat{V}_{D_i}}=\frac{k_B\bar{T}_i}{Ze\bar{B}\bar{L}\bar{V}_A}\frac{\boldsymbol{\hat{b}}\times\boldsymbol{\hat{\nabla}}\hat{P}_i}{\hat{n}_i\hat{B}}.
\end{equation}
In several previous BOUT++ edge turbulence models~\cite{xia2013five,xia2013six}, the magnetic field was normalized to background equilibrium profile $B_0$ instead of a fixed value $\bar{B}$. Consequently, electric field and $E\times B$ drift were written as
\begin{equation}
    \boldsymbol{\hat{E}}=-\frac{\boldsymbol{\hat{\nabla}}\left(B_0\hat{\phi}\right)}{B_0}\simeq -\boldsymbol{\hat{\nabla}}\hat{\phi},\quad
    \boldsymbol{\hat{V}_E}=\frac{\boldsymbol{\hat{b}}\times\boldsymbol{\hat{\nabla}}\left(B_0\hat{\phi}\right)}{B_0}\simeq \boldsymbol{\hat{b}}\times\boldsymbol{\hat{\nabla}}\hat{\phi}
\end{equation}
which require $k/L_B\ll 1$ (here $L_B$ is the characteristic length of background magnetic field). This condition is normally satisfied for drift-type turbulence at the tokamak edge. Nevertheless, the new normalization removes this constraint and attains simpler electric field and $E\times B$ drift expressions.

The relation between parallel current $J_\parallel$ and parallel velocities $V_{\parallel e,i}$ is modified as
\begin{equation}\label{eq:vpe_raw_norm}
    \hat{V}_{\parallel e}=\hat{V}_{\parallel i}-\frac{\bar{B}}{\mu_0\bar{L}\bar{n}\bar{V}_A e}\frac{\hat{J}_\parallel}{\hat{n}_e}.
\end{equation}

To keep the parametric dependence of resistivity and ion viscosity, collision times are also normalized to $\bar{t}$, and the resulting dimensionless $\hat{\tau}_{\alpha\beta}=\tau_{\alpha\beta}^0\hat{T}_\alpha^{3/2}/\hat{n}_\alpha$ where the constant coefficients (independence of $n,T$ evolution but still radially dependent as the Coulomb logarithm $\lambda=\ln \Lambda$ is evaluated based on initial equilibrium) are
\begin{equation}
    \tau_{ee}^0=4.8816\times10^{11}\frac{\bar{T}_e^{3/2}}{\lambda\bar{n}\bar{t}},\quad
    \tau_{ei}^0=3.4518\times10^{11}\frac{\bar{T}_e^{3/2}}{Z\lambda\bar{n}\bar{t}},\quad \tau_{ii}^0=2.0918\times10^{13}\frac{\sqrt{\mu}\bar{T}_i^{3/2}}{Z^4\lambda\bar{n}\bar{t}}.
\end{equation}
The normalized (leading) ion viscosity coefficient then is
\begin{equation}
    \eta_i^0=0.96\hat{n}_i\hat{T}_i\tau\hat{\tau}_{ii}=\epsilon_g \hat{T}_i^{5/2}\quad\text{where~~}\epsilon_g=0.96\tau_{ii}^0\tau_i,
\end{equation}
and the normalized coefficient for ion heating (heat exchange) term is
\begin{equation}
    \frac{2m_e}{m_i\hat{\tau}_{ei}}=\epsilon_q\frac{\hat{n}_e}{\hat{T}_e^{3/2}}\quad\text{where~~}\epsilon_q=\frac{2m_e}{m_i}\frac{1}{\tau_{ei}^0}.
\end{equation}
Similarly, the normalized Spitzer-H\"arm resistivity,
\begin{equation}
    \eta_\parallel=\frac{\eta_\parallel^0}{\hat{T}_e^{3/2}}\quad\text{where~~}\eta_\parallel^0=1.28\times10^{-4}F(Z)\frac{Z\lambda}{\bar{T}_e^{3/2}}\frac{1}{\mu_0\bar{L}\bar{V}_A}.
\end{equation}
Since the electron inertia ($d_e^2$) term is small and mainly affects shear Alfv\'en wave, electron skin depth is assumed to be a (spatial varying) constant while the dependence of density evolution is neglected. The normalized electron skin depth square is
\begin{equation}
    \hat{d}_e^2=\frac{c^2}{\omega_{pe}\bar{L}^2}=\frac{2.8196\times 10^{13}}{\hat{n}_e\bar{n}\bar{L}^2}.
\end{equation}
To further simplify equation expressions, define
\begin{align}
    \alpha_{di}=\frac{k_B\bar{T}_i}{Ze\bar{B}\bar{L}\bar{V}_A},\quad \alpha_{de}=\frac{k_B\bar{T}_e}{e\bar{B}\bar{L}\bar{V}_A}=\frac{\alpha_{di}}{Z\tau_i},\quad \epsilon_p=\frac{k_B\bar{T}_e}{m_i\bar{V}_A^2},\quad \epsilon_j=\frac{\bar{B}}{\mu_0 e\bar{n}\bar{L}\bar{V}_A},\quad \epsilon_v=\frac{m_i}{Ze\bar{B}\bar{t}}.
\end{align}
Here $\alpha_{di},\alpha_{de}$ are often associated with ion and electron diamagnetic drift related terms; while $\epsilon_p$ is associated with curvature drift terms. For simplicity, the over-hat symbol for normalized quantities is omitted for the rest of the paper, and the normalized six-field equations read
\begin{align}
    \label{eq:n_ff}
    \frac{\partial}{\partial t}n_{i} &= -\left(\frac{1}{B}\boldsymbol{\hat{b}}\times\boldsymbol{\nabla}_{\perp}\phi+V_{\parallel i}\boldsymbol{\hat{b}}\right)\cdot\boldsymbol{\nabla} n_{i} -\frac{2n_{i}}{B}\boldsymbol{\hat{b}}\times\boldsymbol{\kappa}\cdot\boldsymbol{\nabla}_{\perp}\phi+\frac{2\alpha_{di}}{ZB}\boldsymbol{\hat{b}}\times\boldsymbol{\kappa}\cdot\boldsymbol{\nabla}_{\perp}P_{e}-n_{i}B\nabla_{\parallel}\left(\frac{V_{\parallel i}}{B}\right)+\frac{\epsilon_j}{Z}B\nabla_\parallel\left(\frac{J_\parallel}{B}\right)+S_n,\\
    \label{eq:w_ff}
    \frac{\partial}{\partial t}\varpi & = -\left(\frac{1}{B}\boldsymbol{\hat{b}}\times\boldsymbol{\nabla}_{\perp}\phi+V_{\parallel i}\boldsymbol{\hat{b}}\right)\cdot\boldsymbol{\nabla}\varpi + B^{2}\boldsymbol{\nabla}_\parallel \left(\frac{J_{\parallel}}{B}\right)+2\epsilon_p\boldsymbol{\hat{b}}\times\boldsymbol{\kappa}\cdot\boldsymbol{\nabla}P -\frac{2}{3}\epsilon_p\boldsymbol{\hat{b}}\times\boldsymbol{\kappa}\cdot\boldsymbol{\nabla}\pi_{ci}\nonumber \nonumber \\
    & -\frac{\alpha_{di}}{2B^2}\left[\boldsymbol{\hat{b}}\times\boldsymbol{\nabla}P_i\cdot\boldsymbol{\nabla}\left(\nabla_{\perp}^{2}\phi\right)\right]+\frac{1}{2}\left(\boldsymbol{\hat{b}}\times\boldsymbol{\nabla}n_{i}\cdot\boldsymbol{\nabla}V_{E}^{2}\right) +\frac{\alpha_{di}}{2B}\left[\boldsymbol{V_E}\cdot\boldsymbol{\nabla}\left(\nabla_{\perp}^{2}P_{i}\right)-\nabla_{\perp}^{2}\left(\boldsymbol{V}_{E}\cdot\boldsymbol{\nabla}P_{i}\right)\right], \\
    \label{eq:vpi_ff}
    \frac{\partial}{\partial t}V_{\parallel i} &= -\left(\frac{1}{B}\boldsymbol{\hat{b}}\times\boldsymbol{\nabla}_{\perp}\phi+V_{\parallel i}\boldsymbol{\hat{b}}\right)\cdot\boldsymbol{\nabla}V_{\parallel i}-\frac{\epsilon_p}{n_i}\left[\nabla_\parallel P +\frac{2}{3}B^{\frac{3}{2}}\nabla_{\parallel}\left(\frac{\pi_{ci}}{B^{\frac{3}{2}}}\right)\right]-\alpha_{di}\frac{\boldsymbol{\hat{b}}}{n_iB}\times\boldsymbol{\nabla}P_i\cdot\boldsymbol{\nabla}V_{\parallel i}-\frac{V_{\parallel i}S_n}{n_i},\\
    \label{eq:apar_ff}
    \frac{\partial}{\partial t}A_\parallel^* &= -\nabla_{\parallel}\phi+\eta_{\parallel}\nabla_{\perp}^{2}A_\parallel+\alpha_{de}\left(\frac{1}{n_{e}}\nabla_{\parallel}P_{e}+0.71\nabla_{\parallel}T_{e}\right), \\
    \label{eq:ti_ff}
    \frac{\partial}{\partial t}T_{i} & = -\left(\frac{1}{B}\boldsymbol{\hat{b}}\times\boldsymbol{\nabla}_{\perp}\phi+V_{\parallel i}\boldsymbol{\hat{b}}\right)\cdot\boldsymbol{\nabla}T_{i} -\frac{2}{3n_{i}}\nabla_{\parallel}q_{\parallel i}  +\epsilon_q\frac{Z^2n_i}{T_e^{3/2}}\left(\frac{T_{e}}{\tau_i}-T_{i}\right) \nonumber \\
    & -\frac{2}{3}T_{i}\left[\left(\frac{2}{B}\boldsymbol{\hat{b}}\times\boldsymbol{\kappa}\right)\cdot\left(\boldsymbol{\nabla}\phi+\alpha_{di}\frac{\boldsymbol{\nabla}P_{i}}{n_{i}}+\frac{5}{2}\alpha_{di}\boldsymbol{\nabla}T_{i}\right)+B\nabla_{\parallel}\left(\frac{V_{\parallel i}}{B}\right)\right] \nonumber \\
    & -\frac{2}{9\tau_i}\frac{\pi_{ci}}{n_{i}}\left[\frac{2}{\sqrt{B}}\nabla_{\parallel}\left(\sqrt{B}V_{\parallel i}\right)-\alpha_{di}\frac{1}{n_{i}B}\boldsymbol{\hat{b}}\cdot\boldsymbol{\nabla}n_{i}\times\boldsymbol{\nabla}T_{i}\right] -\epsilon_v\frac{4}{3B}T_{i}V_{\parallel i}\boldsymbol{\hat{b}}\times\boldsymbol{\kappa}\cdot\boldsymbol{\nabla}V_{\parallel i}+\frac{2 S^E_i}{3n_i}-\frac{T_i S_n}{n_i},\\
    \label{eq:te_ff}
    \frac{\partial}{\partial t}T_{e} & = -\left(\frac{1}{B}\boldsymbol{\hat{b}}\times\boldsymbol{\nabla}_{\perp}\phi+V_{\parallel e}\boldsymbol{\hat{b}}\right)\cdot\boldsymbol{\nabla}T_{e} - \frac{2}{3n_{e}}\nabla_{\parallel}q_{\parallel e} +0.71\frac{2}{3}\epsilon_j\frac{T_{e}}{n_{e}}B\nabla_{\parallel}\left(\frac{J_{\parallel}}{B}\right) -\epsilon_q\frac{Zn_i}{T_e^{3/2}}\left(T_{e}-\tau_iT_{i}\right) \nonumber \\
    & -\frac{2}{3}T_{e}\left[\left(\frac{2}{B}\boldsymbol{\hat{b}}\times\boldsymbol{\kappa}\right)\cdot\left(\boldsymbol{\nabla}\phi-\alpha_{de}\frac{\boldsymbol{\nabla}P_{e}}{n_{e}}-\frac{5}{2}\alpha_{de}\boldsymbol{\nabla}T_{e}\right)+B\nabla_{\parallel}\left(\frac{V_{\parallel e}}{B}\right)\right] +\frac{2}{3\epsilon_p}\frac{1}{n_{e}}\eta_{\parallel}J_{\parallel}^{2}+\frac{2 S^E_e}{3n_e}-\frac{T_e S_n}{n_e},
\end{align}
with
\begin{equation}\label{eq:vort_ff}
    \varpi= \frac{n_i}{B}\left(\nabla_\perp^2\phi +\frac{1}{n_i}\boldsymbol{\nabla}_\perp\phi\cdot\boldsymbol{\nabla_\perp} n_i+\alpha_{di}\frac{\nabla_\perp^2P_i}{n_i}\right),
\end{equation}
\begin{equation}
    A_\parallel^*=\left(1-d_e^2\nabla_\perp^2 \right)A_\parallel
\end{equation}
\begin{equation}\label{eq:pg_ff}
    \pi_{ci}= -\epsilon_gT_i^{5/2}\left[ \frac{\boldsymbol{\hat{b}}\times\boldsymbol{\kappa}}{B}\cdot\left( \boldsymbol{\nabla}\phi+\alpha_{di}\frac{\boldsymbol{\nabla}P_i}{n_i}\right) + \frac{2}{\sqrt{B}}\partial_\parallel(\sqrt{B}V_{\parallel i})\right],
\end{equation}
\begin{equation}
    J_\parallel = J_{\parallel 0} + j_\parallel\quad\text{where}~~ j_\parallel=-\nabla_\perp^2 A_\parallel,
\end{equation}
and
\begin{equation}
    V_{\parallel e}=V_{\parallel i}-\epsilon_j\frac{J_\parallel}{n_e}.
\end{equation}

In addition, to improve the numerical stability artificial forth order hyper-diffusion terms are optional on the right-hand-side of the equations for each quantity. Depending on the nature of the problem and the resolution, these terms could be individually turned on to dissipate grid-scale turbulence eddies if necessary.

\section{Coordinate, mesh, operators and boundary conditions}\label{sec:coordinate}

We briefly outline the coordinate system, mesh, operators and boundary conditions that are employed in the Landau fluid model in this section to lay the foundation for further discussion.

\subsection{Coordinate and mesh}
In the magnetic fusion community, there are more than a dozen choices of coordinate systems tailored for different research problems~\cite{sauter2013tokamak}; while in BOUT++, there are three coordinate systems are involved at different stages, namely the cylindrical coordinate system $(R,\phi,Z)$, the flux coordinate system $(\psi,\theta,\zeta)$ and the field-aligned coordinate $(x,y,z)$. 
The cylindrical coordinate system $(R,\phi,Z)$ is preferred by experimentalists due to its simplicity representing plasma equilibrium; the flux coordinate system $(\psi,\theta,\zeta)$ is widely used by theorists in analytical analysis; and the field-aligned coordinate $(x,y,z)$ is employed in the actual BOUT++ simulation due to its superior numerical efficiency.
Because the cylindrical coordinate system $(R,\phi,Z)$ is mainly involved in the equilibrium calculation and mesh generation (e.g., tokamak grid generator HYPNOTOAD~\cite{dudson2015bout++}) in BOUT++, here we focus on the latter two coordinate systems.

Before diving into the details, we need to clarify our notations. In BOUT++ notation, $(R,\phi,Z)$ is always a right-handed coordinate system as the one defined in EFIT~\cite{lao1985reconstruction} and kinetic EFIT. $(\psi,\theta,\zeta)$ and $(x,y,z)$, however, could be either right-handed or left-handed, depending on the poloidal field $B_\theta$ direction. The positive toroidal $\zeta$ direction is defined as counter-clockwise when viewing from the top of torus, i.e., the opposite of the ``normal" direction considered by experimentalists. The positive poloidal $\theta$ direction is defined as the direction of the polodial magnetic field $\gv{B}_\theta$ produced by a positive toroidal plasma current $I_p$. In experiments, plasma current $I_p$ can be negative or positive, thus, the $B_\theta$ in our notation, can also be negative or positive.

In the flux coordinate system, the tangent-basis (covariant-basis) vectors are $\gv{e}_i=\partial \gv{r}/\partial i$ and the reciprocal-basis (contravariant-basis) vectors are $\gv{e}^i=\nabla i$ with $i=\psi,\theta,\zeta$. The corresponding scale factor $h_i=|\gv{e}_i|$ are
\begin{equation}
|\gv{e}_\psi|=\frac{1}{R|B_\theta|}, \quad |\gv{e}_\theta|=h_\theta, \quad |\gv{e}_\zeta|=R,
\end{equation}
and hence
\begin{equation}
|\gv{e}^\psi|=R|B_\theta|, \quad |\gv{e}^\theta|=\frac{1}{h_\theta}, \quad |\gv{e}^\zeta|=\frac{1}{R}.
\end{equation}

The total magnetic field in a tokamak is composed by the poloidal and toroidal magnetic fields; conveniently, it can be written as
\begin{equation}
\gv{B}=\gv{B}_\theta+\gv{B}_\zeta=\nabla \zeta\times \nabla \psi + F(\psi)\nabla\zeta
\end{equation}
with $\psi$ the poloidal flux and $F=RB_\zeta$. The Jacobian of flux coordinate $(\psi,\theta,\zeta)$ is thus simply
\begin{equation}\label{eq:jptz}
J^{-1}=\nabla\psi\times\nabla\theta\cdot\nabla\zeta=\nabla\zeta\times\nabla\psi\cdot\nabla\theta=\gv{B}_\theta\cdot\nabla\theta=B_\theta/h_\theta
\end{equation}
which again can be either positive or negative depending on the sign of poloidal field $\sigma_{B_\theta}=B_\theta/|B_\theta|$, or the direction of plasma current $I_p$.

Although the orthogonal flux coordinate $(\psi, \theta, \zeta)$ works well in analytical analysis, BOUT++ employs a field-aligned coordinate $(x, y, z)$ to take the advantage of the strongly antiseptic feature of magnetized plasma by minimizing the resolution requirement in the field-line direction:
\begin{equation}\label{eq:field-align-coord}
x=\sigma_{B_\theta}(\psi-\psi_0), \quad y=\theta, \quad z=\sigma_{B_\theta}\left(\zeta-\int_{\theta_0}^{\theta} \nu(\psi,\theta)  d \theta \right).
\end{equation}
Here $\psi_0, \theta_0$ are the reference poloidal flux and poloidal angle, normally set to be at the seperatrix and outboard midplane, respectively. The local pitch of the magnetic field $\nu$ is defined as 
\begin{equation}\label{eq:localshear}
\nu=\frac{B_\zeta h_\theta}{B_\theta R}=\frac{(F/R)h_\theta}{B_\theta R}=FJ/R^2.
\end{equation}

With Equation~(\ref{eq:field-align-coord}), one can write down the transformation of the contravariant and covaraint basis ($\gv{e}^i, \gv{e}_i$) between two coordinate systems,
\begin{equation}
\begin{pmatrix}
\gv{e}^x \\ \gv{e}^y \\ \gv{e}^z 
\end{pmatrix}
=
\begin{pmatrix}
\sigma_{B_\theta} & 0 & 0 \\ 0 & 1 & 0 \\ -\sigma_{B_\theta}I & -\sigma_{B_\theta}\nu & \sigma_{B_\theta}
\end{pmatrix}
\begin{pmatrix}
\gv{e}^\psi \\ \gv{e}^\theta \\ \gv{e}^\zeta 
\end{pmatrix}
, \quad
\begin{pmatrix}
\gv{e}_x \\ \gv{e}_y \\ \gv{e}_z 
\end{pmatrix}
=
\begin{pmatrix}
\sigma_{B_\theta} & 0 & \sigma_{B_\theta}I \\ 0 & 1 & \nu \\ 0 & 0 & \sigma_{B_\theta}
\end{pmatrix}
\begin{pmatrix}
\gv{e}_\psi \\ \gv{e}_\theta \\ \gv{e}_\zeta 
\end{pmatrix}
,
\end{equation}
where the integrated local shear $I(\psi,\theta)$ is defined as
\begin{equation}\label{eq:intshear}
I(\psi,\theta)=\int_{\theta_0}^{\theta}\frac{\partial\nu(\psi,\theta)}{\partial\psi}d\theta.
\end{equation}

As
\begin{equation}\label{eq:exyz2eptz}
\nabla x=\sigma_{B_\theta}\nabla \psi, \quad \nabla y=\nabla \theta, \quad \nabla z=\sigma_{B_\theta} \left( \nabla \zeta-I
\nabla \psi -\nu \nabla \theta\right),
\end{equation}
we therefore yield the contravariant metric tensor $g^{ij}=\gv{e}^i\cdot\gv{e}^j=\nabla i\cdot\nabla j$ and the covariant metric tensor, $g_{ij}=(g^{ij})^{-1}$,
\begin{equation}
g^{ij}=
\begin{pmatrix}
(RB_\theta)^2 & 0 & -I(RB_\theta)^2 \\
0 & 1/h_\theta^2 & -\sigma_{B_\theta}\nu/h_\theta^2 \\
-I(RB_\theta)^2 & -\sigma_{B_\theta}\nu/h_\theta^2 & I^2(RB_\theta)^2 +B^2/(RB_\theta)^2
\end{pmatrix}
, \quad
g_{ij}=
\begin{pmatrix}
\frac{1}{(RB_\theta)^2}+I^2R^2 & \sigma_{B_\theta}\frac{B_\zeta h_\theta IR}{B_\theta} & IR^2 \\
\sigma_{B_\theta}\frac{B_\zeta h_\theta IR}{B_\theta} & \frac{B^2 h_\theta^2}{B_\theta^2} & \sigma_{B_\theta}\frac{B_\zeta h_\theta R}{B_\theta} \\
IR^2 & \sigma_{B_\theta}\frac{B_\zeta h_\theta R}{B_\theta} & R^2
\end{pmatrix}
.
\end{equation}
Similarly, one can write down the cross products of contravariant basis of field-aligned coordinate system,
\begin{equation}\label{eq:xyzcross}
    \nabla x\times\nabla y=(R|B_\theta|/h_\theta)\guv{e}_\zeta, \quad \nabla x\times\nabla z=-\gv{B},\quad \nabla y\times\nabla z=1/(Rh_\theta)\guv{e}_\psi+IR|B_\theta|/h_\theta\guv{e}_\zeta.
\end{equation}

Note that in the field-aligned coordinate system, constant $x$ and $z$ means staying on same magnetic line, as the magnetic field could simply be represented in the Clebsch form
\begin{equation}\label{eq:xyzb}
\gv{B}=\nabla z\times\nabla x = \frac{B_\theta}{h_\theta} \left( \gv{e}_\theta +\nu \gv{e}_\zeta \right) = \frac{B_\theta}{h_\theta} \gv{e}_y.
\end{equation}
The above expression also indicates that the unit vector $\guv{b}=\gv{B}/|B|=B_\theta/(h_\theta |B|) \gv{e}_y$ could be in either the same or the opposite direction of $\gv{e}_y$ depending on the sign of $B_\theta$.
The Jacobian of $(x,y,z)$ is given by
\begin{equation}\label{eq:jxyz}
J^{-1}=\nabla x \times \nabla y \cdot \nabla z=B_\theta/h_\theta,
\end{equation}
which is equivalent to Equation~(\ref{eq:jptz}).

\begin{figure}[ht]
\centering
\includegraphics[width=\linewidth]{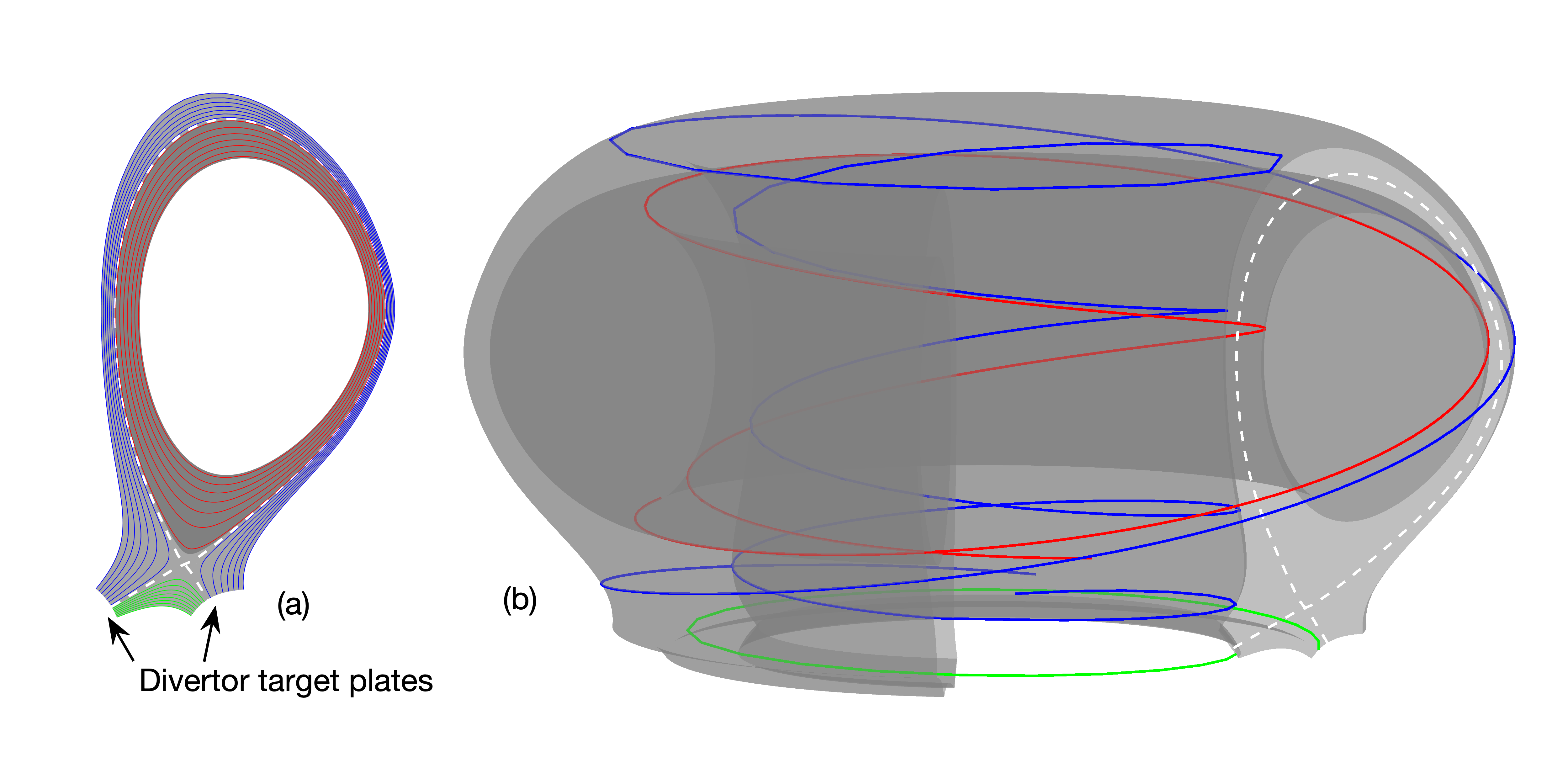}%
\caption{(a) Flux surfaces in 2D poloidal plane, and (b) magnetic field-lines in the 3D torus. White dashed lines denote separatrix, red, blue and green lines represents magnetic field-lines (b) and their poloidal projections (a) in the closed flux region, scrape-off-layer and private flux region respectively. \label{fig:flux_fieldline}}%
\end{figure}
 
\begin{figure}[ht]
\centering
\includegraphics[width=\linewidth]{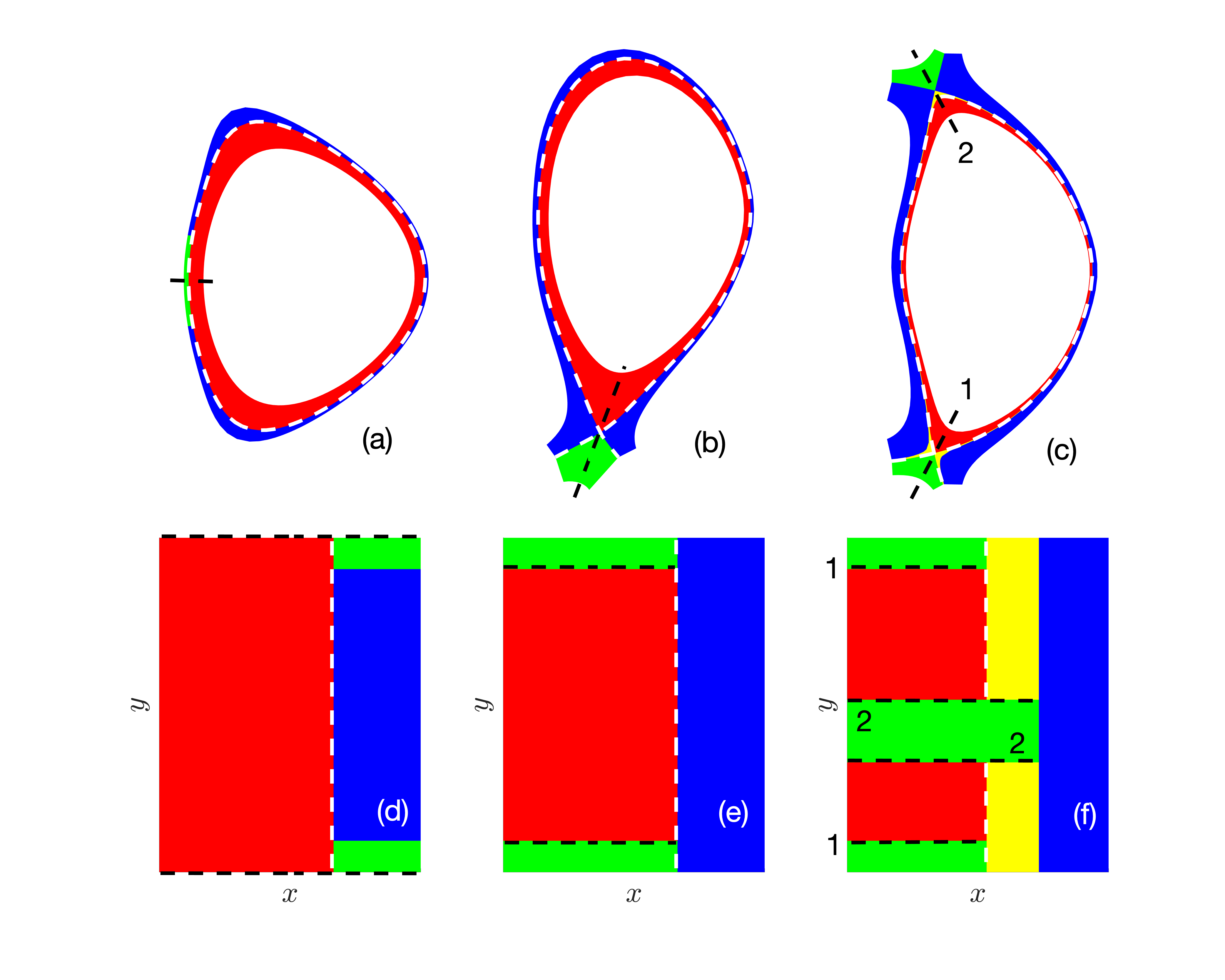}%
\caption{Typical poloidal planes for (a) limiter, (b) single null and (c) double null divertor configurations and their corresponding $xy$ meshes in the index space. White dashed lines denote last-closed flux surface, or separatrix. Red, blue (and yellow), and green areas represent closed flux region, scrape-off-layer and private flux region. Black dashed lines indicate the locations of branch-cut. \label{fig:bout_mesh}}%
\end{figure}

As mentioned previously there are two types of magnetic field lines in the tokamak edge region. Figure~\ref{fig:flux_fieldline} illustrates the flux surfaces and the magnetic field lines at the tokamak edge region in a lower single null divertor configuration, the most common tokamak discharge nowadays. Inside the separatrix (white dashed), the projections of magnetic field-lines on the poloidal plane form the so-called closed flux surfaces (red lines); while outside the separatrix, the projections yield open flux surfaces as the magnetic field-lines are terminated at the divertor target plates. Depending on whether the magnetic field-lines pass through the top poloidal plane, the open flux region is further categorized into the scrape-off-layer (blue lines) and the private flux region (green lines).
With these three somewhat detached regions, generating a mesh that could be parallelized efficiently is no longer straightforward. However, by carefully dividing the poloidal plane into separated blocks and connecting them based on their locations in configuration space, the simulation domain remains topologically rectangular in the index space as shown in Figure~\ref{fig:bout_mesh}.
To truthfully capture the geometric information, BOUT++'s tokamak meshes are normally based on ideal MHD equilibrium either attained from external solvers such as CORSICA or GATO, or reconstructed from experimental measurements via EFIT or kinetic EFIT. Auxiliary mesh generation codes (e.g., HYPERNOTOAD~\cite{dudson2015bout++}) are developed to produce high accuracy BOUT++ meshes with desired resolution and domain. Therefore, BOUT++ is capable to simulate realistic tokamak discharges in limiter, single-null and double-null divertor configurations.

The authors remark that despite the discussion in this paper emphasizes on tokamak plasma, in principle, the generic representation of magnetic field configuration in BOUT++ makes the Landau fluid model a flexible tool to study plasmas in a variety kinds of magnetic configurations, for example, the linear devices~\cite{popovich2010analysis}, the helical plasmas and even the stellarator plasmas~\cite{shanahan2018fluid}.

\subsection{Operators}

Here we list the most common operations involved in the fluid simulations generalized in the field-aligned coordinate system. In BOUT++, a vector $\gv{v}$ is by default in covariant form unless it is declared otherwise. However, it may be transformed in contravariant form during certain operations to reduce the overall computation. 

\subsubsection{Dot, cross,  gradient, divergent, curl and Laplacian operators}

The dot and cross products of two vectors $\gv{u}$ and $\gv{v}$ can be calculated either in covariant or contravariant forms,
\begin{align}
\gv{u}\cdot\gv{v}&=\sum_{i,j}g_{ij}u^i v^j=\sum_{i,j} g^{ij}u_i v_j, \\
\gv{u}\times\gv{v}&=J\sum_k(u^iv^j-u^jv^i)\gv{e}^k=\frac{1}{J}\sum_k(u_iv_j-u_jv_i)\gv{e}_k.
\end{align}
The gradient of a scalar $f$ is most convenient to be evaluated in covariant components form ($u_i$) as
\begin{equation}
\gv{u}=\nabla f=\sum_i \frac{\partial f}{\partial i}\gv{e}^i.
\end{equation}
In some cases, only the perpendicular component of $\nabla f$ is needed which is defined as
\begin{equation}
\gv{u}_\perp= \nabla_\perp f= -\guv{b}\times(\guv{b}\times \nabla f)=\frac{-1}{B^2}\gv{B}\times(\gv{B}\times \nabla f) .
\end{equation}
The divergence of a vector $\gv{u}$ is most convenient to be evaluated in contravariant components form ($u^i$) as
\begin{equation}
f=\nabla \cdot \gv{u} = \frac{1}{J}\sum_i \frac{\partial}{\partial i}(J u^i).
\end{equation}
The curl of a vector $\gv{v}$ is again most convenient to be evaluated in covariant components form ($v_i$) and the result is in contravariant components form ($u^i$),
\begin{equation}
\gv{u}=\nabla\times\gv{v}=\frac{1}{J}\sum_k \left(\frac{\partial v_j}{\partial i}-\frac{\partial v_i}{\partial j}\right)\gv{e}_k
=\frac{1}{J} \left(\frac{\partial v_z}{\partial y}-\frac{\partial v_y}{\partial z}\right)\gv{e}_x
+\frac{1}{J} \left(\frac{\partial v_x}{\partial z}-\frac{\partial v_z}{\partial x}\right)\gv{e}_y
+\frac{1}{J} \left(\frac{\partial v_y}{\partial x}-\frac{\partial v_x}{\partial y}\right)\gv{e}_z.
\end{equation}
The Laplacian operator, by definition, is
\begin{equation}
\nabla^2f=\frac{1}{J}\sum_i\frac{\partial}{\partial i}\left(J \sum_j \frac{\partial f}{\partial j}\gv{e}^j\cdot \gv{e}^i \right),
\end{equation}
or, in explicit form 
\begin{equation}
\begin{split}
\nabla^2f&= |\nabla x|^2 \frac{\partial^2 f}{\partial x^2}+ |\nabla y|^2 \frac{\partial^2 f}{\partial y^2}+ |\nabla z|^2 \frac{\partial^2 f}{\partial z^2}+2\nabla x\cdot \nabla y \frac{\partial^2 f}{\partial x \partial y}+2\nabla x\cdot \nabla z \frac{\partial^2 f}{\partial x \partial z}+2\nabla y\cdot \nabla z \frac{\partial^2 f}{\partial y \partial z}\\
&+\nabla^2 x\frac{\partial f}{\partial x}+\nabla^2 y\frac{\partial f}{\partial y}+\nabla^2 z\frac{\partial f}{\partial z}.
\end{split}
\end{equation}

\subsubsection{Poisson bracket, advection, parallel derivatives and curvature operators}
In the fluid model, an important type of operator is called ``Poisson bracket" defined as
\begin{equation}
[f,g]=\frac{\guv{b}}{B}\times \nabla f\cdot\nabla g=\frac{\guv{b}}{B}\cdot \nabla f\times \nabla g,
\end{equation}
which physically means convection processes (e.g., $E\times B$ convection if $f=\phi$, or magnetic flutter effects if $f=A_\parallel$).
Rewriting $\nabla f\times \nabla g$ as
\begin{equation}
\nabla f\times \nabla g = \left(\frac{\partial f}{\partial x}\frac{\partial g}{\partial y}-\frac{\partial f}{\partial y}\frac{\partial g}{\partial x}\right)\gv{e}^x\times\gv{e}^y +\left(\frac{\partial f}{\partial x}\frac{\partial g}{\partial z}-\frac{\partial f}{\partial z}\frac{\partial g}{\partial x}\right)\gv{e}^x\times\gv{e}^z + \left(\frac{\partial f}{\partial y}\frac{\partial g}{\partial z}-\frac{\partial f}{\partial z}\frac{\partial g}{\partial y}\right)\gv{e}^y\times\gv{e}^z,
\end{equation}
and with the aid of Equations~(\ref{eq:xyzcross}) and (\ref{eq:xyzb}), the full expression for Poisson bracket becomes
\begin{equation}
\gv{B}\times \nabla f\cdot\nabla g=B_\zeta\frac{R|B_\theta|}{h_\theta}\left(\frac{\partial f}{\partial x}\frac{\partial g}{\partial y}-\frac{\partial f}{\partial y}\frac{\partial g}{\partial x}\right) -B^2\left(\frac{\partial f}{\partial x}\frac{\partial g}{\partial z}-\frac{\partial f}{\partial z}\frac{\partial g}{\partial x}\right) + IB_\zeta\frac{R|B_\theta|}{h_\theta} \left(\frac{\partial f}{\partial y}\frac{\partial g}{\partial z}-\frac{\partial f}{\partial z}\frac{\partial g}{\partial y}\right).
\end{equation}

Because $E\times B$ drift $\gv{v}_E=\gv{E}\times\gv{B}/B^2=\gv{B}\times\nabla \phi/B^2$, $E\times B$ advection then becomes
\begin{equation}\label{eq:exbconv}
\gv{v}_E\cdot \nabla g=\frac{1}{B^2}\gv{B}\times\nabla \phi \cdot \nabla g = [\phi,g].
\end{equation}

In general, an advection process of quantity $f$ with velocity $\gv{u}$ is easier to be evaluated in contravariant form as
\begin{equation}
\begin{split}
\gv{u}\cdot\nabla f&=u_x\frac{\partial f}{\partial x} |\nabla x|^2+u_y\frac{\partial f}{\partial y} |\nabla y|^2+u_z\frac{\partial f}{\partial z} |\nabla z|^2 + \left( u_x\frac{\partial f}{\partial y}+u_y\frac{\partial f}{\partial x}\right) \nabla x \cdot \nabla y\\
&\left( u_x\frac{\partial f}{\partial z}+u_z\frac{\partial f}{\partial x}\right) \nabla x \cdot \nabla z+ \left( u_y\frac{\partial f}{\partial z} +u_z\frac{\partial f}{\partial y}\right) \nabla y \cdot \nabla z~\text{(in covariant form)}\\
&=u^x\frac{\partial f}{\partial x} +u^y\frac{\partial f}{\partial y} +u^z\frac{\partial f}{\partial z}~\text{(in contravariant form)}
\end{split}
\end{equation}

The parallel derivative along an \textit{unperturbed} magnetic field $\guv{b}=\gv{B}/B$ is
\begin{equation}
\nabla_\parallel^{(0)} f =\guv{b}\cdot \nabla f=\frac{1}{B}\left( \nabla z\times \nabla x\right)\cdot\left( \nabla y \frac{\partial f}{\partial y}\right)=\frac{1}{JB}\frac{\partial f}{\partial y}=\frac{B_{\theta}}{h_\theta B}\frac{\partial f}{\partial y},
\end{equation}
and the double derivative then becomes
\begin{equation}
\nabla_\parallel^{2(0)} f=\frac{B_{\theta}}{h_\theta B}\frac{\partial}{\partial y}\left(\frac{B_{\theta}}{h_\theta B}\frac{\partial f}{\partial y}\right)=\frac{B_{\theta}}{h_\theta B}\frac{\partial}{\partial y}\left(\frac{B_{\theta}}{h_\theta B}\right)\frac{\partial f}{\partial y}+\frac{B_{\theta}^2}{h_\theta^2 B^2}\frac{\partial^2 f}{\partial y^2}.
\end{equation}
The derivative along the normalized \textit{perturbed} magnetic field $\tilde{\gv{b}}=\tilde{\gv{B}}/B=(1/B)\nabla A_\parallel \times \guv{b}$ is
\begin{equation}
\tilde{\gv{b}}\cdot \nabla f=(1/B^2)\nabla A_\parallel \times \gv{B}\cdot \nabla f=-(1/B^2)\gv{B} \times \nabla A_\parallel  \cdot \nabla f=-[A_\parallel,f].
\end{equation}


The last type of important operators in the toroidal system is the curvature operator $\guv{b}\times\guv{\kappa}\cdot\nabla f$, where $\guv{\kappa}$ is the magnetic line curvature, defined as $\guv{\kappa}=\guv{b}\cdot\nabla\guv{b}=-\guv{b}\times(\nabla\times\guv{b})$.
The curvature operator could be rewritten in a similar form of the advection operator
\begin{equation}
\guv{b}\times\guv{\kappa}\cdot\nabla f=v_x\frac{\partial f}{\partial x}+v_y\frac{\partial f}{\partial y}+v_z\frac{\partial f}{\partial z}
\end{equation}
where the velocity components $v_i=\guv{b}\times\guv{\kappa}\cdot \nabla i$ with $i=x,y,z$ are calculated during mesh generation and then passed along as the prescribed inputs. Neglecting the perturbed magnetic field contribution is justified because normally it is three or more orders of magnitude smaller than the equilibrium magnetic field.

It should be pointed out that in order to avoid numerical error accumulation in a sheared mesh (e.g., tokamak edge), radial derivative in BOUT++ in fact is often not done in field-aligned coordinate system but ``shifted"~\cite{umansky2009status} and evaluated in a so-called ``quasi-ballooning" coordinate system. The implementation and advantage of shifted radial derivative is thoroughly discussed in References
~\cite{dudson2009bout++,seto2019interplay}.

\subsection{Boundary conditions}

For a toroidal plasma system like tokamak, it is natural to impose periodic boundary condition in the toroidal ($z$) direction. In the radial direction ($\psi$ for sheared mesh, otherwise, $x$), a large variety of boundary options is available within BOUT++ for different physics models, including the most commonly used Dirichlet, Neumann, Robin boundary conditions as well as the ``relaxed" boundary which evolves the quantity towards the given boundary conditions at certain rate in order to prevents transient discontinuities occurring at the boundaries. In general, for the global Landau fluid model, if applicable, Dirichlet boundary is applied to the ``perturbed", non-axisymmetric (i.e., $n\neq 0$) quantities, and Neumann boundary is applied to the ``equilibrium" axisymmetric (i.e., $n=0$) quantities to minimize the fluctuations at the boundary.
The parallel $y$ direction has two distinct boundary conditions depending on its corresponding magnetic topology. As illustrated in Figure~\ref{fig:flux_fieldline}, inside the separatrix magnetic field-lines wind around the major axis while simultaneously swirl around minor (magnetic) axis, and nest so-called closed flux surfaces. In most cases, a magnetic field-line will not end up at the same toroidal location after finishing one poloidal turn; therefore, a twist-shift operation~\cite{xu2008boundary,umansky2009status} is used to ensure the parallel continuity in the closed flux region. 
In the open field-line region, i.e., the scrape-off-layer and the private flux region classical Bohm sheath criteria~\cite{stangeby2000plasma} are enforced at the boundary where magnetic field-lines are intercepted by limiter or divertor target plates,
\begin{align}
    V_{\parallel i} &=\pm \max[C_s,|V_{\parallel i}|],\\
    J_\parallel&=\pm n_e e \left[ C_s - \frac{v_{th,e}}{2\sqrt{\pi}}\exp\left(-\frac{e\Phi}{k_BT_e}\right)\right],\\
    \label{eq:qsheath}
    Q_{\parallel e,i}&=\pm \gamma_{e,i}n_{e,i}k_Bv_{th e,i}T_{e,i},
\end{align}
Here $C_s$ is the local sound speed, arising from the presheath electrostatic potential. In divertor configuration, theory and model predict that ion could be supersonic near the divertor plates due to the Laval nozzle effect~\cite{cohen1999drifts,togo2019self}, supersonic ion flow is also occasionally observed in UEDGE and BOUT++ transport simulations~\cite{supersonic}. Therefore, ion parallel velocity $V_{\parallel i}$ is compared to the the local sound speed $C_s$ at each time step -- if it is less than $C_s$, then $|V_{\parallel i}|= C_s$; otherwise, it is free evolving, i.e., supersonic flow is allowed.
The electron/ion sheath energy transmission factors $\gamma_{e,i}$ depend on secondary emission coefficient, local electron and ion temperature ratio, target plate floating potential~\cite{stangeby1984plasma}. The typical values used in BOUT++ simulations are $\gamma_e\simeq 7, \gamma_i\simeq 2.5$. As will be discussed in next section, the implementation of this boundary condition depends on the specific parallel heat flux model to be used.
In addition the following boundary conditions on density, vorticity and electrostatic potential are applied to eliminate the potential build-up of large discontinuities at the boundary,  
\begin{equation}
    \nabla_\parallel n_i=\nabla_\parallel \varpi =0,
\end{equation}
\begin{equation}
    \nabla_\parallel\phi=-\eta_\parallel j_\parallel + \frac{\nabla_\parallel P_e}{en_e}+\frac{0.71k_B}{e}\nabla_\parallel T_e.
\end{equation}

\section{Parallel heat flux models}\label{sec:lf}

\subsection{Braginskii (Spitzer-H\"arm) heat flux model}
The Braginskii transport model~\cite{braginskii1965transport} derived under the assumption that plasma is magnetized and collsional requires that $L_\perp\gg \rho$ and $L_\parallel\gg \lambda_\text{mfp}$. Here $\rho$ is the particle (electron and ion) gyroradius, $\lambda_\text{mfp}$ is the particle mean free path, and $L_\perp,L_\parallel$ are the characteristic lengths in perpendicular and parallel directions. Under the strong collisional limit, the well-known Braginskii (or, Spitzer-H\"arm) heat flux model is deduced by assuming the perturbed particle distribution slightly deviates from the equilibrium Maxwellian distribution,
\begin{equation}\label{eq:q_Brag}
    q_{\parallel j}^\text{B}=-\kappa_\parallel^j\nabla_\parallel \left(k_B T_j\right) \quad \text{with~}\kappa_\parallel^e=3.2\frac{n_ek_BT_e\tau_{ei}}{m_e},\quad \kappa_\parallel^i=3.9\frac{n_ik_BT_i\tau_{ii}}{m_i}
\end{equation}
which implies that heat flux depends on local temperature gradient, and heat (or, energy) is transported solely through collisions (or, random walk process) among particles.

Recent particle simulations indicate that Braginskii parallel heat flux model is reasonably accurate (i.e., less than 10\% error) when the parallel inhomogeneity length is at least on the two order of magnitude longer than the particle mean free path (i.e., $L_\parallel/\lambda_\text{mfp}\geq O(10^2)$) and becomes invalid when $L_\parallel/\lambda_\text{mfp}\leq O(10)$~\cite{guo2014parallel}.

In the tokamak edge region, $L_\perp$ is the equilibrium length such as $L_n, L_T$ which are typically a few centimeters, while $L_\parallel\sim 2\pi q R$ in large aspect ratio limit is on the order of tens meters where $q\approx 4\sim 6$ is the safety factor and $R$ is the major radius of the machine.
Assuming that plasma temperature is isotropic (i.e., $T_\perp=T_\parallel=T$), then $\rho=v_\text{th}/\omega_{c}\propto m^{1/2}T^{1/2}$ and $\lambda_\text{mfp}\sim v_\text{th}\tau \propto T^2/n$ is independent of mass. As modern and future tokamaks produce higher and higher temperature plasmas, the latter requirement ($L_\parallel\gg \lambda_\text{mfp}$) is no long be satisfied due to the strong temperature dependence of $\lambda_\text{mfp}$. 
For example, Table~\ref{tab:brag} examines the validity of Braginskii model for the H-mode edge of three tokamaks: Alcator C-Mod (compact high field, relatively high collisionality machine), DIII-D (medium size, relatively low collisionality machine) and ITER (future experimental reactor). 
These estimate indicate that Braginksii parallel heat flux model merely holds near the separatrix and breaks down at the pedestal region.
The direct consequence of violation of strong collisionality criterion $L_\parallel \gg \lambda_\text{mfp}$ is that the classical Braginskii (or, Spitzer-H\"arm) parallel heat flux model (Equation~\ref{eq:q_Brag}) is no longer accurate and overestimates the parallel heat flux by orders of magnitude in the weakly collisional regime as demonstrated in Figure~\ref{fig:closure_comp}.
Although the classical Braginskii parallel heat flux model is still widely used for more collisional L-mode edge plasma simulations nowadays,~\cite{zhu2018up,zholobenko2020thermal} a more advanced means of parallel heat flux evaluation for weakly collsional plasma is needed in fluid simulations.

\begin{table}[h]
    \centering
    \caption{Braginskii validation for Deuterium plasma}
    \label{tab:brag}
    \begin{tabular}{|c|c|c|c|c|c|c|}
    \hline
         &  \multicolumn{3}{|c|}{Mid-pedestal} & \multicolumn{3}{|c|}{Separatrix} \\ \hline
         & CMod & DIII-D & ITER & CMod & DIII-D & ITER \\ \hline
    $B(T)$ & 5 & 2 & 5 & 5 & 2 & 5 \\ \hline
    $q$    & 4 & 4 & 3 & 5 & 5 & 4 \\ \hline
    $R(m)$ & 0.68 & 1.67 & 6.2 & 0.68 & 1.67 & 6.2 \\ \hline
    $n(10^{20}m^{-3})$ & 1.6 & 0.6 & 0.6 & 0.8 & 0.4 & 0.3 \\ \hline
    $T(eV)$ & 150 & 250 & 2000 & 50 & 80 & 200 \\ \hline
    $L_\perp(cm)$ & 1 & 1.5 & 3 & 2 & 3 & 6 \\ \hline
    $L_\parallel(m)$ & 17 & 42 & 117 & 21 & 52 & 156 \\ \hline
    $\rho_e (cm)$ & $5.83\times 10^{-4}$ & $1.88\times10^{-3}$ & $2.13\times10^{-3}$ & $3.37\times10^{-4}$ & $1.06\times10^{-4}$ & $6.73\times10^{-4}$ \\ \hline
    $\rho_i (cm)$ & $3.53\times10^{-2}$ & $1.14\times10^{-1}$ & $1.29\times10^{-1}$ & $2.04\times10^{-2}$ & $6.45\times10^{-2}$ & $4.08\times10^{-2}$ \\ \hline
    $\lambda_\text{mfp} (m)$ & 1.4 & 9.5 & 505 & 0.34 & 1.6 & 12.1 \\ \hline
    \end{tabular}
\end{table}

\subsection{Flux-limited model}
Intuitively, in a collisionless plasma, parallel heat transport is mainly done by particles freely moving along the magnetic field-lines which also sets a upper limit of heat flux. In fluid description, this free streaming heat flux is defined as $q_{\parallel}^\text{FS}=nk_BTv_{th}$ where $v_{th}$ is the thermal speed. Hence, a flux-limited expression for parallel heat flux is constructed to constrain the unrealistically large value predicted by Braginskii model~\cite{omotani2013non},
\begin{equation}\label{eq:q_FL}
    q_\parallel^\text{FL}=\left( \frac{1}{q_{\parallel}^\text{B}} +\frac{1}{\alpha q_{\parallel}^\text{FS}}\right)^{-1}
\end{equation}
where $\alpha\in[0.1,1]$ is a parameter to be adjusted in different scenarios.
However, Equation (\ref{eq:q_FL}) becomes singular at where $q_{\parallel}^\text{B}+\alpha q_{\parallel}^\text{FS}\approx 0$ as by definition, $q_{\parallel}^\text{FS}$ is always positive while $q_{\parallel}^\text{B}$ can be negative depending on local temperature gradient and $|q_{\parallel}^\text{B}|\gg q_{\parallel}^\text{FS}$ in the weakly collisional regime.
In practice, $q_{\parallel}^\text{FS}$ is replaced with the ``diffusive" free streaming heat flux $q_{\parallel}^\text{FS(d)}=-nv_{th}L_\parallel\nabla_\parallel (k_B T)$ which ensures it has the same sign as $q_{\parallel}^\text{B}$ at all locations. Consequently, the resulting flux-limited parallel heat flux is again dependent on local temperature gradient, similar to the Braginskii model. It can be rewritten as $q_{\parallel j}^\text{FL}=-\kappa_\text{eff}^j\nabla_\parallel (k_B T_j)$
with the effective thermal conductivity
\begin{equation} 
    \kappa_\text{eff}^j=\frac{\alpha n_j v_{th,j} L_\parallel \kappa_\parallel^j}{\alpha n_j v_{th,j} L_\parallel + \kappa_\parallel^j}.
\end{equation}

\subsection{Landau fluid closure}
\begin{figure}[ht]
\centering
\includegraphics[width=0.5\linewidth]{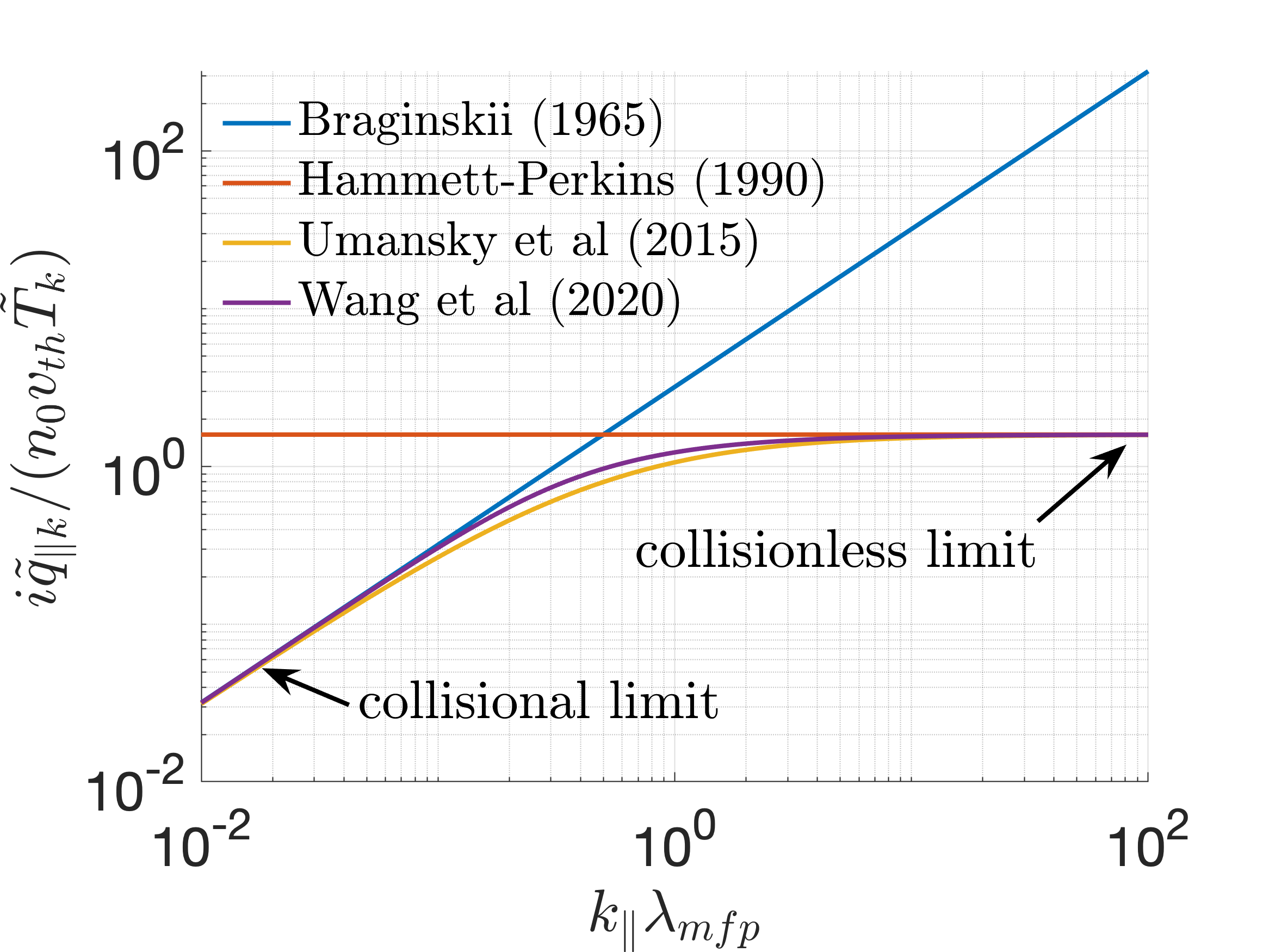}%
\caption{Parallel heat flux v.s. collisionality. \label{fig:closure_comp}}%
\end{figure}

Although the flux-limited parallel heat flux model overcomes the overestimation of Braginskii model, it is still an ad hoc model.
A more rigorous approach is to develop a theory-based parallel heat flux model, for example, the Landau fluid model for weakly collisional plasmas~\cite{umansky2015modeling}
\begin{equation}\label{eq:q_LF}
    q_{\parallel j,k}^\text{LF}=
    -n_{j}k_B\sqrt{\frac{8}{\pi}}v_{th,j}\frac{ik_\parallel}{|k_\parallel|+\alpha_j/\lambda_j}T_{j,k},
\end{equation}
where $\alpha_e\simeq 0.499$ and $\alpha_i\simeq 0.409$. This Landau fluid closure is similar to the closure by incorporating the Krook collision operator~\cite{beer1996toroidal,snyder1997landau} with 
slightly different coefficients and an extension of Hammett-Perkins closure~\cite{hammett1990fluid} -- the first and the simplest Landau fluid closure that describes plasma kinetic responses in the collsionless limit, 
\begin{equation}\label{eq:q_HP}
    q_{\parallel,k}^\text{HP}=-n_{j}k_B\sqrt{\frac{8}{\pi}}v_{th,j}\frac{ik_\parallel}{|k_\parallel|}T_{j,k}
\end{equation}
As a result, it seamlessly bridges collsional Braginskii expression (\ref{eq:q_Brag}) and collsionless Hammett-Perkins closure (\ref{eq:q_HP}) as shown in Figure~\ref{fig:closure_comp}. Comparing with the exact kinetic closures with collisional effects~\cite{wang2020deep}, this closure predicts a slightly lower heat flux (with a maximum difference by 25$\%$) in the intermediate collisionality regime.

\begin{figure}[ht]
\centering
\includegraphics[width=0.5\linewidth]{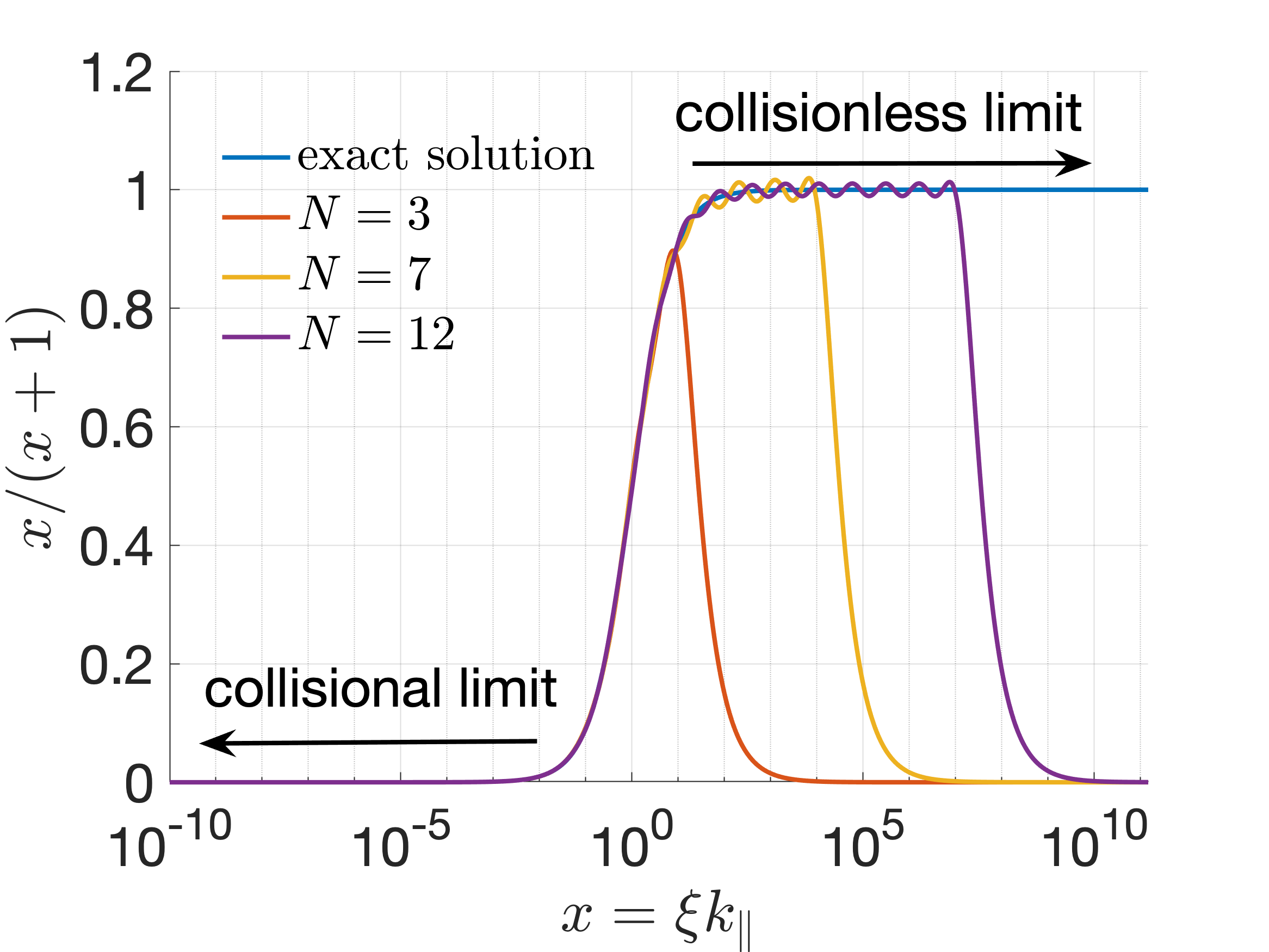}%
\caption{Accuracy and the validity of the Landau fluid closure v.s. collisionality for different numbers of Lorentizians. \label{fig:qfit}}%
\end{figure}

Implementing such a closure in BOUT++ turbulence model is numerically challenging.
This is because Landau fluid closures are usually entail wave-vector space (i.e., $k-$space) and implicitly assume periodicity along magnetic field-lines; while BOUT++ turbulence model calculates quantities in configuration space. A novel solution to resolve these issues is the fast non-Fourier method~\cite{dimits2014fast} which approximates parallel heat flux with summation of Lorentizians as
\begin{equation}\label{eq:sum_lorentizian}
    q_{\parallel j,k}\approx \sum_{n=0}^{N-1} q_{\parallel j}^{(n)},
\end{equation}
where
\begin{equation}
    q_{\parallel j}^{(n)}=
    -n_{j}k_B\sqrt{\frac{8}{\pi}}v_{th,j}\cfrac{\alpha_n \xi_j k_\parallel}{(\xi_j k_\parallel)^2+\beta_n^2}i T_{j,k}.
\end{equation}
Here $\xi_j=\lambda_{mfp,j}/\alpha_j$ and $(\alpha_n,\beta_n)$ are optimized fitting coefficients of
\begin{equation}
    \frac{x}{x+1}\approx \sum_0^{N-1} \frac{\alpha_n x}{x^2+\beta_n^2}
\end{equation}
as the ones given in Table 2 of Reference~\cite{chen2019extension}.
It is therefore straightforward to transform the above Lorentizians back to configuration space by replacing $ik_\parallel$ and $k_\parallel^2$ with $\nabla_\parallel$ and $-\nabla_\parallel^2$, 
\begin{equation}\label{eq:lorentizian}
    \left(\beta_n^2-\xi_j^2\nabla_\parallel^2\right)q_{\parallel j}^{(n)}=-n_{j}k_B\sqrt{\frac{8}{\pi}}v_{th,j} \xi_j\alpha_n\nabla_\parallel T_j,
\end{equation}
and the evaluation of parallel heat flux in $k-$space becomes solving a series of boundary value problems in configuration space along unperturbed magnetic field-lines.
The fast non-Fourier method has been developed and validated with a one-dimensional uniform grid~\cite{chen2019extension}. As illustrated in Figure~\ref{fig:qfit}, it attains more accurate approximation and reaches further into the collisionless regime by adding more Lorentizians. In most tokamak edge simulations, $N=7$ is sufficient as it is valid for $\xi k_\parallel \sim k_\parallel \lambda_{mfp}<10^4$ with a relative error about 2\%. One can further push the applicable limit to $k_\parallel \lambda_{mfp}<10^7$ and/or reduce the relative error to about 1\% by using $N=12$ Lorentizians.
This treatment of Landau fluid closure has now been extended and implemented in BOUT++ transport and turbulence codes to deal with three-dimensional divertor geometry.

\begin{figure}[ht]
\centering
\includegraphics[width=\textwidth]{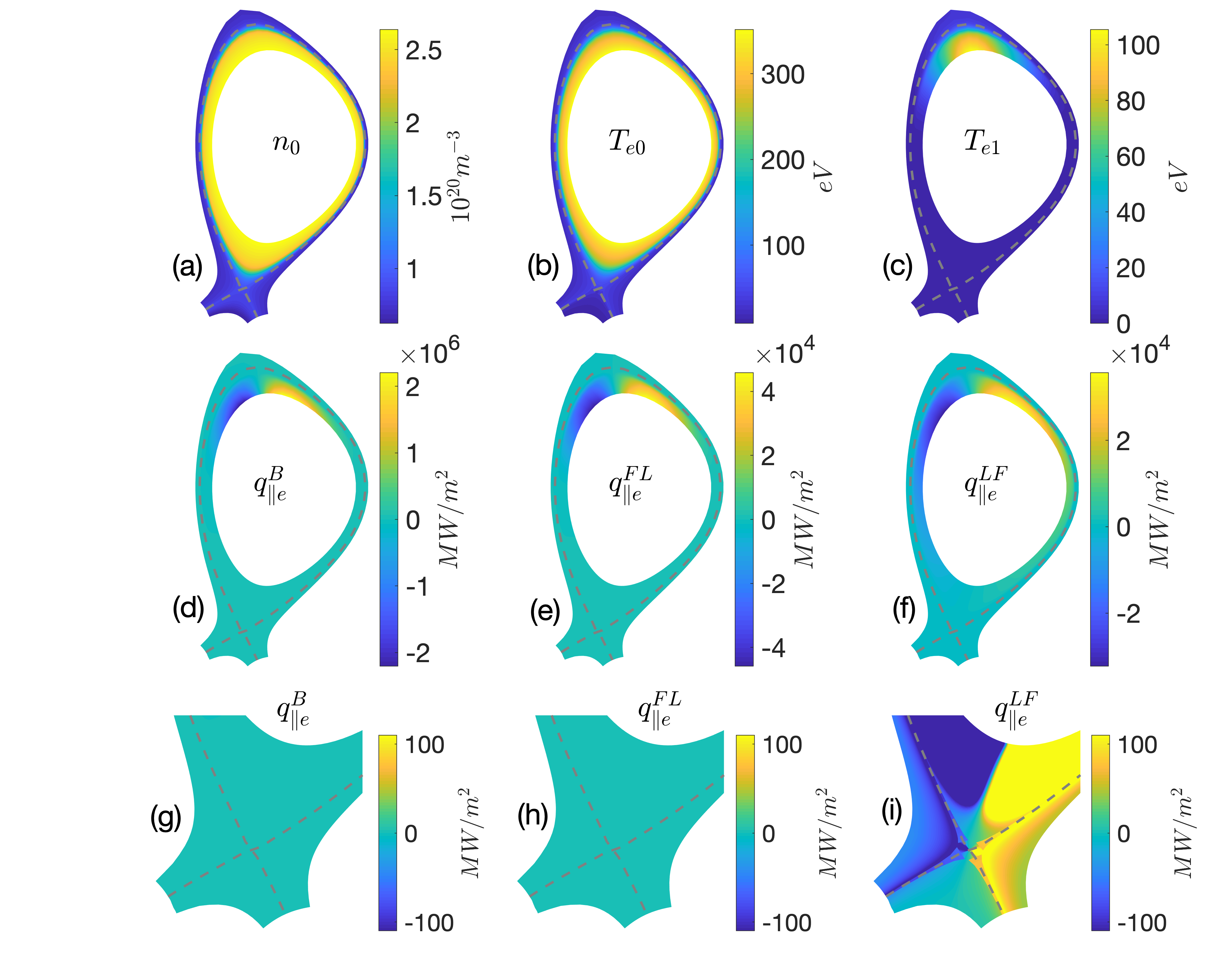}
\caption{Comparison of parallel heat flux for typical C-Mod H-mode parameters: singly charged deuterium plasma equilibrium (a) density $n_0$, (b) electron temperature $T_{e0}$, (c) perturbed electron temperature $T_{e1}$ and the parallel heat flux predictions from (d) Braginskii model, (e) flux-limited model ($\alpha=0.1$), (f) Landau fluid closure as well as their zoom-in plots at the divertor region.}
\label{fig:cmod_qpar}
\end{figure}

\begin{figure}[ht]
\centering
\includegraphics[width=\textwidth]{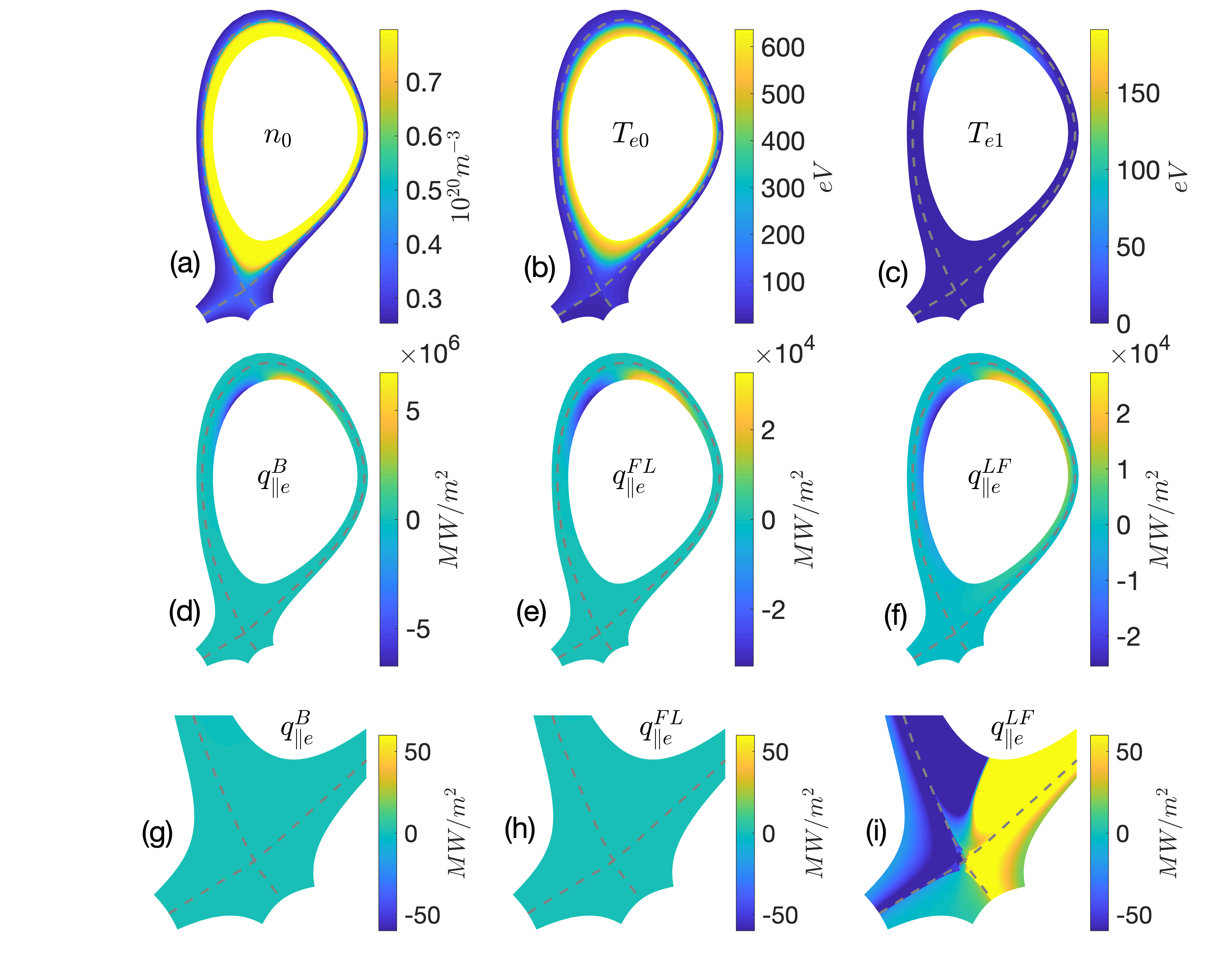}
\caption{Comparison of parallel heat flux for typical DIII-D H-mode parameters: singly charged deuterium plasma equilibrium (a) density $n_0$, (b) electron temperature $T_{e0}$, (c) perturbed electron temperature $T_{e1}$ and the parallel heat flux predictions from (d) Braginskii model, (e) flux-limited model ($\alpha=0.5$), (f) Landau fluid closure as well as their zoom-in plots at the divertor region.}
\label{fig:d3d_qpar}
\end{figure}

To highlight the discrepancies between the original Braginskii heat flux model, flux-limited model and the Landau fluid closure in a realistic tokamak geometry, two scenarios are considered -- a relatively collisional C-Mod discharge (Figure~\ref{fig:cmod_qpar}) and a less collisional DIII-D discharge (Figure~\ref{fig:d3d_qpar}) based on their typical H-mode parameters. In both scenarios, electron parallel heat flux is evaluated assuming there is a relatively large (30\%) temperature perturbation is on the top portion of the equilibrium profiles (e.g., to imitate vertical displacement event-type instability).

The first key observation is that Braginskii model predicts a parallel heat flux only at the tokamak top region where temperature gradient is finite (thus, ``local” model) and its amplitude is about two orders of magnitude larger than either flux-limited or the Landau fluid estimate.
The flux-limited model seems able to effectively scale down the heat flux amplitude close to that predicted by the Landau fluid closure (Figures~\ref{fig:cmod_qpar}(e) versus (f), Figures~\ref{fig:d3d_qpar}(e) versus (f)); nevertheless, it is still local.
On the other hand, the Landau fluid closure obviously exhibits the ``nonlocal" effect as the predicted parallel heat flux extends to the region where local temperature gradient vanishes (e.g., the tokamak bottom region). More interestingly, our tests also demonstrate the transition from local-dominant transport to nonlocal-dominant transport. In C-Mod plasma, parallel heat flux still marginally depends on the local temperature gradient (Figures~\ref{fig:cmod_qpar}(f)). As the plasma becomes less collisional, this local effect weakens and the nonlocal effect dominates as expected (Figures~\ref{fig:cmod_qpar}(f) versus~\ref{fig:d3d_qpar}(f)).

Among the three parallel heat flux models, only the Landau fluid closure is able to consistently enforce the ambipolar sheath heat flux boundary condition $q_\parallel^\text{sheath}$ at the divertor targets (i.e., Equation~\ref{eq:qsheath}) as shown in Figures~\ref{fig:cmod_qpar}(i) and~\ref{fig:d3d_qpar}(i). This is again due to the distinct inherent transport characteristics. Because Braginskii and flux-limited models are local, $q_\parallel^\text{sheath}$ is actually not enforced on parallel heat flux but on the plasma temperature as $\nabla_\parallel T=\pm q_\parallel^\text{sheath}/\kappa$ which tends to produce a steep parallel gradient at the divertor targets. While for Landau fluid closure, $q_\parallel^\text{sheath}$ is provided directly as the Dirichlet boundary condition and hence results in a smoother parallel heat flux profile.

\begin{figure}[ht]
\centering
\includegraphics[width=\textwidth]{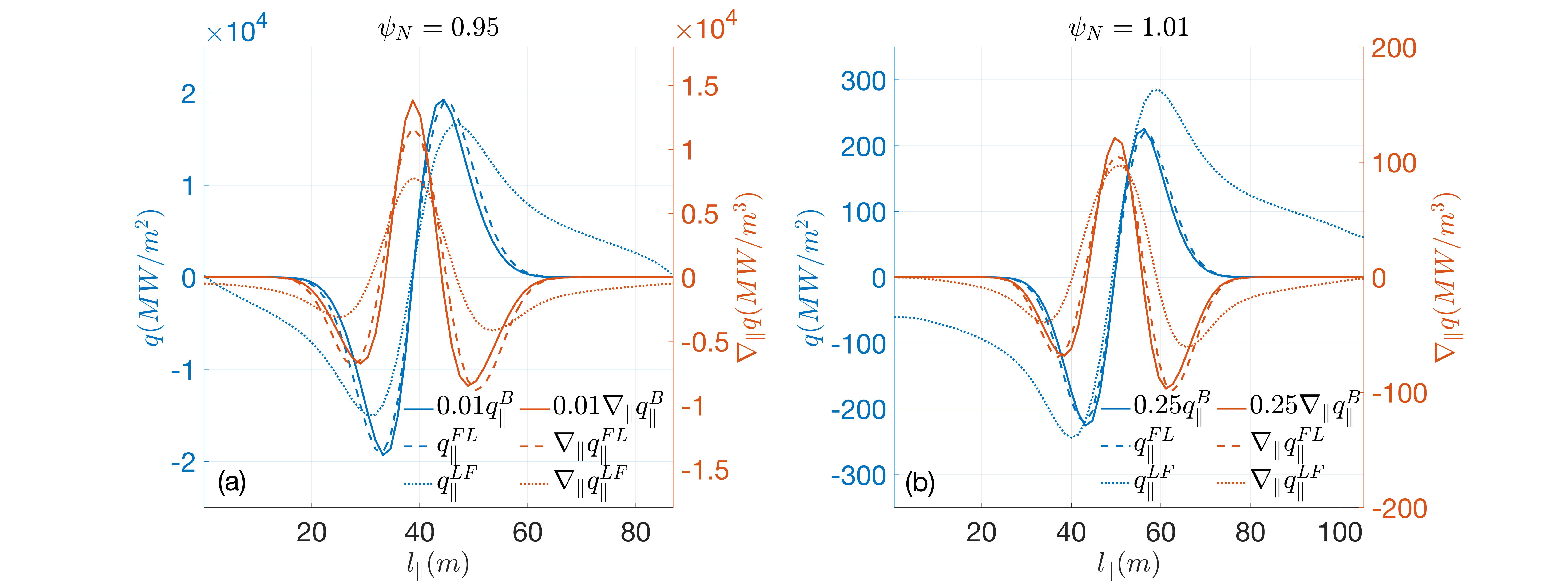}
\caption{$q_\parallel$ and $\nabla_\parallel q_\parallel$ predicted by three models along field-lines (a) at pedestal top ($\psi_N=0.95$) and (b) in the SOL ($\psi_N=1.01$) in the DIII-D test (i.e., Figure~\ref{fig:d3d_qpar}).}
\label{fig:d3d_qpar_comp}
\end{figure}

Given the dramatic discrepancies between three parallel heat flux models, one would expect that plasma parallel dynamics may also be altered significantly. Perhaps a more important quantity is the parallel gradient of parallel heat flux ($\nabla_\parallel q_\parallel$) since it directly influences temperature evolution.
Our initial analysis suggests that Landau fluid closure results in a nearly one order of magnitude weaker heat conduction comparing to the value given by the flux-limited model both in the closed flux region and in the SOL,
as attested by the substantial diminishing amplitudes of $\nabla_\parallel q_\parallel$ (red dotted lines) in Figure~\ref{fig:d3d_qpar_comp}. 
However, the impact of the improved Landau fluid closure on edge instability, turbulence, transport and most importantly, divertor heat flux width~\cite{xu2019simulations} doesn't fit in the scope of this paper and will be left for a future study.

Although the Landau fluid closure is arguably the most sophisticated treatment of parallel heat flux in nowadays fluid edge turbulence models, it is worth to clarify that there are still a few subtleties in this approach. For instance, the temperature anisotropy effect is neglected in order to be consistent with Braginskii transport equations, and the classical Bohm's sheath criteria for collisional plasma is used at the boundaries in the open field-line region. The former can be addressed in a more complete BOUT++ gyro-Landau-fluid turbulence model.
Although one may argue that this Bohm's sheath boundary condition is only valid in collisional fluid regime, the mean kinetic (from the half-Maxwellian) and fluid velocities are actually very similar ($\bar v=\sqrt{8T/m_i\pi}$ vs $c_s$) because ions that reach the divertor plate will escape. While the heat flux predicted by Landau fluid closure (\ref{eq:q_LF}) is found to have a good agreement with the result from a Fokker-Planck code with a small amplitude perturbation in weakly collisional regime~\cite{umansky2015modeling}, a further validation with extended parameter regime will be performed and reported in future publications.

\subsection{Normalization}
For the sake of completeness, the normalization of different parallel heat flux models are listed below. In BOUT++ transport and turbulence codes, $q_{\parallel j}$ is naturally normalized with  $\bar{q}_j=\bar{n}k_B\bar{V}_A\bar{T}_j$ and the normalized parallel heat flux for Braginskii expression and flux-limited model along with their thermal conductivities are
\begin{equation}
    q_{\parallel j}^\text{B}=-\kappa_\parallel^j \nabla_\parallel T_j\text{~and~} q_{\parallel j}^\text{FL}=-\kappa_\text{eff}^j \nabla_\parallel T_j
\end{equation}
with
\begin{equation}
    \kappa_\parallel^e=3.2\tau_{ei}^0T_e^{3/2}v_{th,e}^2\ , \quad \kappa_\parallel^e=3.9\tau_{ii}^0T_i^{3/2}v_{th,i}^2\ , \quad
    \kappa_\text{eff}^j=\frac{\alpha n_j v_{th,j} L_\parallel \kappa_\parallel^j}{\alpha n_j v_{th,j} L_\parallel + \kappa_\parallel^j}.
\end{equation}
Here the plasma density and temperature $n_j, T_j$, thermal speed $v_{th,j}$ and parallel characteristic length $L_\parallel$ are normalized to $\bar{n}$, $\bar{T}_j$, $\bar{V}_A$ and $\bar{L}$ respectively.

For the Landau fluid closure, Equation~(\ref{eq:lorentizian}) that solves Lorentizians of parallel heat flux after normalization becomes
\begin{equation}
    \left(\beta_n^2-\xi_j^2\nabla_\parallel^2\right)q_{\parallel j}^{(n)}=-n_{j}\sqrt{\frac{8}{\pi}}v_{th,j} \xi_j\alpha_n\nabla_\parallel T_j,
\end{equation}
where $n_j$, $T_j$, $\xi_j$ and $v_{th,j}$  are again normalized to $\bar{n}$, $\bar{T}_j$, $\bar{L}$ and $\bar{V}_A$.

\section{Field solver}\label{sec:zonal}
In the drift-reduced fluid formalism, Laplacian inversion is required to evaluate electrostatic potential $\phi$ from general vorticity $\varpi$, and parallel magnetic potential $A_\parallel$ (or perturbed magnetic flux $\psi{=}{-}A_\parallel$) from modified magnetic flux $A_\parallel^*$ if electron inertia is included in the model,
\begin{align}
    \label{eq:laplace2d}
    \nabla_\perp^2\phi+\frac{1}{n_i}\nabla_\perp\phi\cdot\nabla_\perp n_i&=\frac{B\varpi}{n_i}-\frac{\alpha_{di}}{n_i}\nabla_\perp^2P_i, \\
    d_e^2\nabla_\perp^2 A_\parallel-A_\parallel&=-A_\parallel^*.
\end{align}
Therefore, Laplacian inversion solvers are designed in BOUT++ to solve both equations in a generalized form
\begin{equation}\label{eq:laplaceequ}
    d\nabla_\perp^2 f +\frac{1}{c}\nabla_\perp c\cdot \nabla_\perp f+af=b.
\end{equation}

Since BOUT++ employs the field aligned coordinate system, the perpendicular component of Laplacian operator used in the vorticity equation and Ohm's law, by definition, is
\begin{align}\label{eq:delp2}
    \nabla_\perp^2 f&= (\nabla^2-\nabla_\parallel^2) f \\ \nonumber
    &= G_x\frac{\partial f}{\partial x}+G_y\frac{\partial f}{\partial y}+G_z\frac{\partial f}{\partial z}+g^{xx}\frac{\partial^2f}{\partial x^2}+g^{yy}\frac{\partial^2 f}{\partial y^2}+g^{zz}\frac{\partial^2 f}{\partial z^2} \\ \nonumber
    &+2g^{xy}\frac{\partial^2 f}{\partial x\partial y}+2g^{xz}\frac{\partial^2 f}{\partial x\partial z}+2g^{yz}\frac{\partial^2 f}{\partial y\partial z}-\frac{1}{g_{yy}}\frac{\partial^2 f}{\partial y^2}-\frac{1}{J}\frac{\partial}{\partial y}\left(\frac{J}{g_{yy}}\right)\frac{\partial f}{\partial y}
\end{align}
where $G_i=\nabla^2i$ for $i=x,y,z$ and the $\partial^2_{xy}$ term is omitted as $g^{xy}=0$. 
Instead of directly solving in the matrix inversion problem within 3D configuration space, BOUT++ separates the unknown quantify $f$ into non-axisymmetric ($n\neq0$) and axisymmetric ($n=0$) components to improve the numerical performance by reducing the dimensionality of the problem. Here $n$ represents the toroidal mode number.

\subsection{Non-axisymmetric ($n\neq 0$) field}
For the non-axisymmetric field, considering the strongly anisotropic nature of magnetized plasmas, the parallel derivatives are normally several orders of magnitude smaller than the perpendicular derivatives, i.e., $k_\parallel{\ll} k_\perp$. Therefore we may neglect any $\partial/\partial y$ terms and also the $G_x, G_z$ terms (as the grid is normally uniform in $x$ and $z$ direction), and Equation~(\ref{eq:delp2}) becomes
\begin{equation}
\begin{split}
\nabla_\perp^2f &= g^{xx}\frac{\partial^2f}{\partial x^2}+g^{zz}\frac{\partial^2 f}{\partial z^2}
    +2g^{xz}\frac{\partial^2 f}{\partial x\partial z} \\
    &= (RB_\theta)^2\left\lbrace\frac{\partial^2 f}{\partial x^2}-2I \frac{\partial^2 f}{\partial x \partial z}+\left[I^2+\frac{B^2}{(RB_\theta)^4} \right]\frac{\partial^2 f}{\partial z^2} \right\rbrace.
\end{split}
\end{equation}
Rewriting above equation in flux coordinate system to drop all the $I$ terms which may introduce large error in a sheared mesh, and then applying Fourier transform along the toroidal direction, yields a 1D boundary value problem and can be solved efficiently with the tridiagonal solvers~\cite{seto2019interplay}.

\subsection{Axisymmetric ($n=0$) field}
However, the above approach doesn't apply to the axisymmetric $k_z=0$ (or $n=0$) zonal field component. Intuitively, this means that the compressed 1D profile ($x$ distribution) does not contain enough information to reconstruct a 2D structure ($xy-$plane); therefore, one needs to include $y$ variation in order to correctly solving the zonal field component. In other similar edge turbulence codes, this process is normally done with a hard-coded algebraic multigrid solver~\cite{halpern2016gbs,francisquez2019multigrid,zholobenko2021electric}. BOUT++ utilizes external libraries (e.g., PETSc~\cite{petsc-web-page}, HYPRE~\cite{hypre-web-page}) to solve this problem thus has a more flexible option on the numerical algorithm to be used~\cite{dudson2017hermes,seto2019interplay}. 
In BOUT++ 2D transport code, the relaxation method is used to solve the zonal potential equation for a steady-state solution.~\cite{li2018calculation}

For $n=0$ mode, $\partial/\partial z$ terms vanishes, and Equation~(\ref{eq:delp2}) becomes 
\begin{equation}
    \nabla_\perp^2 f = G_x\frac{\partial f}{\partial x}+G_y\frac{\partial f}{\partial y}+g^{xx}\frac{\partial^2f}{\partial x^2}+g^{yy}\frac{\partial^2 f}{\partial y^2}
    -\frac{1}{g_{yy}}\frac{\partial^2 f}{\partial y^2}-\frac{1}{J}\frac{\partial}{\partial y}\left(\frac{J}{g_{yy}}\right)\frac{\partial f}{\partial y}.
\end{equation}
Above equation is then discretized either based on the second order finite difference scheme (thus, 5-point stencil) or the forth order finite difference scheme (thus, 9-point stencil) and solved as a 2D boundary value problem.

\begin{figure}[!ht]
\centering
\includegraphics[width=0.8\linewidth]{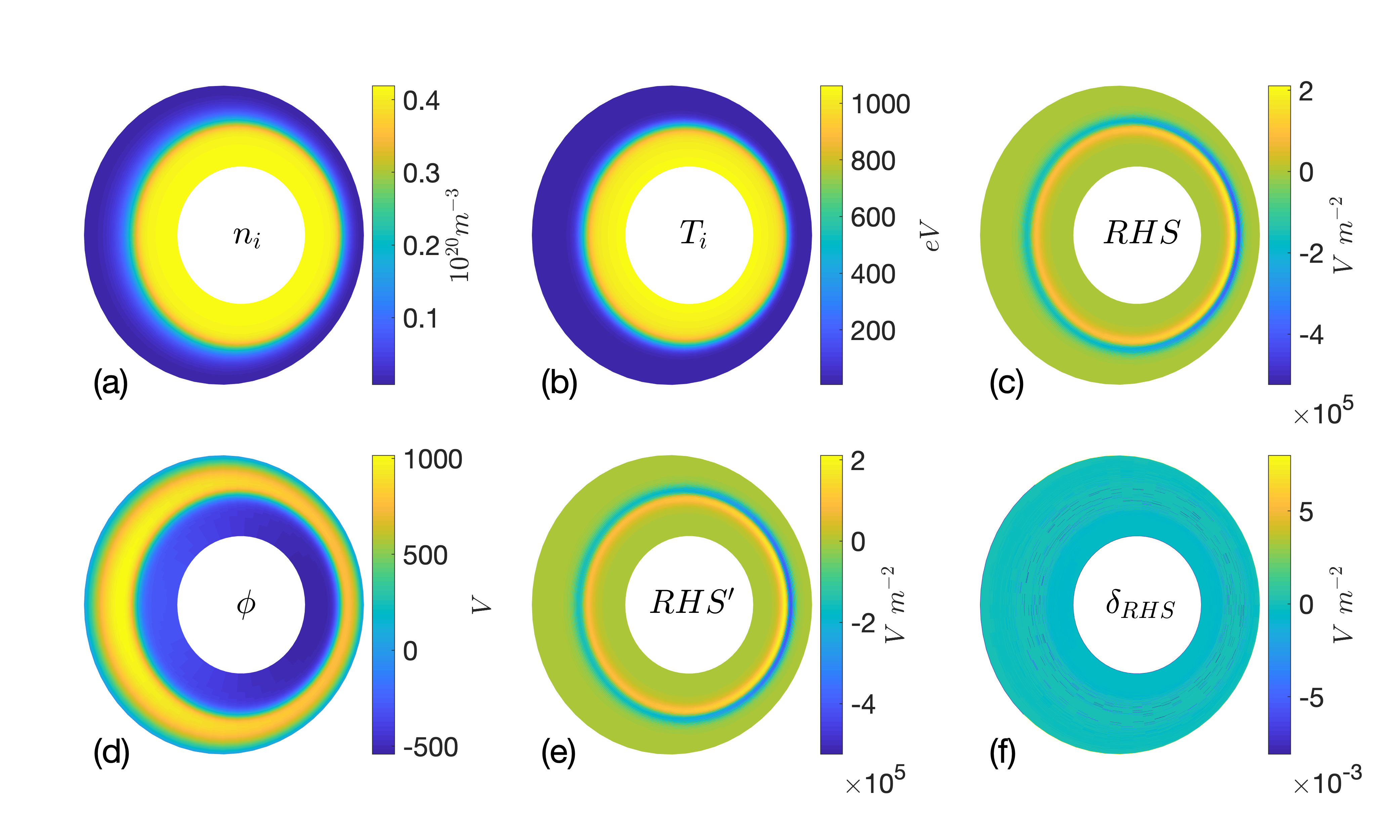}%
\caption{Inverse-and-forward electrostatic potential test of the axisymmetric solver in a shift-circular configuration: (a) density, (b) ion temperature and (c) evaluated $RHS$ profiles, (d) calculated $\phi$, (e) re-evaluated $RHS'$ and (f) the absolute error $\delta_{RHS}$. This test is done with cross term and $\epsilon_R=10^{-10}$. \label{fig:cbm_zonal}}%
\end{figure}

To verify that the solver is implemented properly, an inverse-and-forward test on electrostatic potential is performed. That is, giving an self-consistent set of profiles ($n_i,T_i,B$) and assuming $\varpi=0$, evaluate the right-hand-side of Equation~(\ref{eq:laplace2d}), i.e., 
\begin{equation}
    \nabla_\perp^2\phi+\frac{1}{n_i}\nabla_\perp\phi\cdot\nabla_\perp n_i=-\frac{\alpha_{di}}{n_i}\nabla_\perp^2 P_i=RHS;
\end{equation} 
and calculate $\phi$ by solving the 2D boundary value problem; then re-evaluate the left-hand-side of Equation~(\ref{eq:laplace2d}) (denoted as $RHS'$) based on $\phi$; and finally the absolute error $\delta_{RHS}=RHS'-RHS$ is used to quantify the performance of the solver in this quasi-realistic setting. In this test, Equation~(\ref{eq:laplace2d}) is discretized with a 5-point stencil, and the generalized minimal residual (GMRES) method and the successive over-relaxation (SOR) method in the PETSc library are picked as the matrix solver and the preconditioner. Dirichlet boundary condition ($\phi=\text{constant}$) is applied on the outside boundary (wall-side) and Neumann boundary condition ($\partial \phi/\partial \psi=0\propto V_{E,p}$) is enforced on the inside boundary (core-side) to eliminate any external poloidal angular momentum sourcing. Additional options, such as the absolute and relative error tolerances $\epsilon_A,\epsilon_R$ set the iteration criteria, could be used to adjust the overall solver accuracy.
Figure~\ref{fig:cbm_zonal} shows a typical test in a shift-circular configuration with resolution $(n_x,n_y)=(516,64)$. The absolute error $\delta_{RHS}\sim 10^{-2}$ (Figure~\ref{fig:cbm_zonal}(f)) is about eight orders of magnitude smaller than the initial input (Figure~\ref{fig:cbm_zonal}(c)), suggesting that the axisymmetric field solver is correctly implemented. Moreover, the resulting error $\delta_{RHS}$ tends to collocate at where $RHS$ is finite.

\begin{figure}[!ht]
\centering
\includegraphics[width=\linewidth]{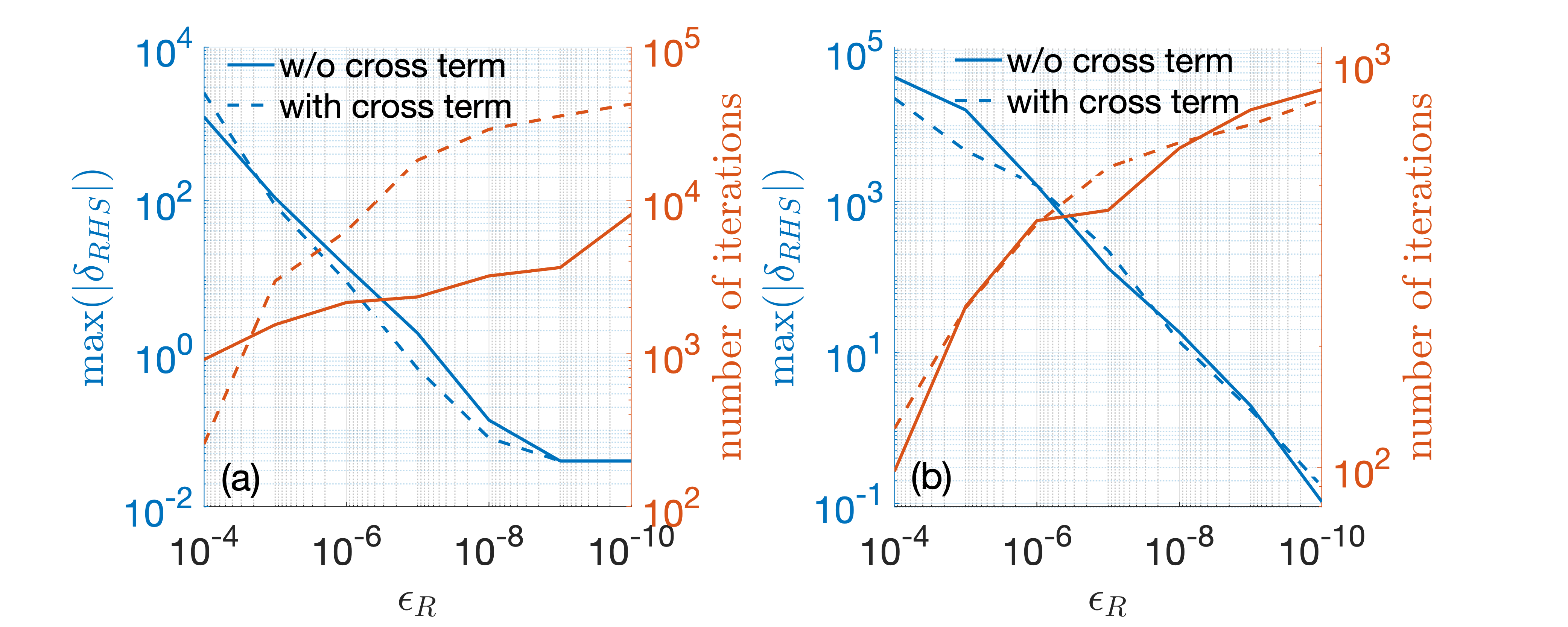}%
\caption{The maximum absolute error (blue lines) and the number of iterations (red lines) in the inverse-and-forward tests v.s. relative tolerance $\epsilon_R$ with (dashed lines) and without (solid lines) cross term with (a) shift circular and (b) lower single null configuration. \label{fig:laplace2d_conv}}%
\end{figure}

A convergence test is carried out by varying the relative error tolerance $\epsilon_R$ with and without the $\nabla_\perp n_i \cdot\nabla_\perp \phi$ cross term on the left-hand-side of Equation~(\ref{eq:laplace2d}), and the result is summarized in Figure~\ref{fig:laplace2d_conv}(a).
Not surprisingly, the axisymmetric field solver converges. As $\epsilon_R$ decreases, the maximum absolute error of the test $|\delta_{RHS}|_{max}$ decreases linearly as well till it saturates at $10^{-2}$ level.
As expected, the solver slows down as $\epsilon_R$ becomes smaller, i.e., the number of iterations required to achieve certain criterion increases.
However, the apparent discrepancy between the two red lines indicates that the number of iterations is reduced substantially when the cross term is neglected (i.e., applying Boussinessq approximation -- a widely used but not fully justified assumption in reduced-fluid research) especially with a fixed relatively small $\epsilon_R$.
It is worth to point out that the exact number of iterations depends on quite a few factors, for example, grid resolution, smoothness and consistency of the profiles ($RHS$ and $n_i$), choices of matrix solver and preconditioner and so on.

\begin{figure}[ht]
\centering
\includegraphics[width=0.8\linewidth]{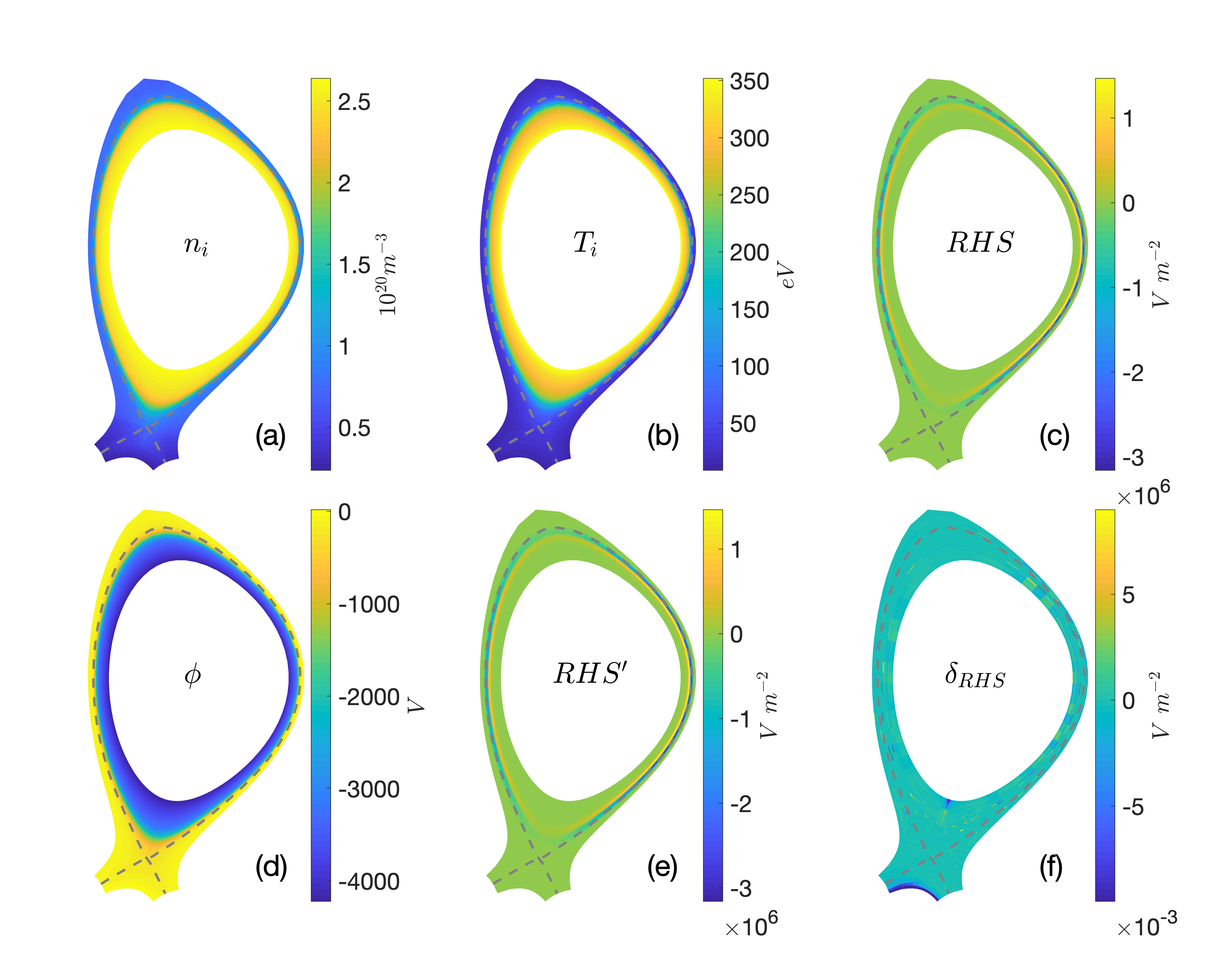}%
\caption{Inverse-and-forward electrostatic potential test of the axisymmetric solver in a C-Mod-like lower single null divertor configuration: (a) density, (b) ion temperature and (c) evaluated $RHS$ profiles, (d) calculated $\phi$, (e) re-evaluated $RHS'$ and (f) the absolute error $\delta_{RHS}$. This test is done with cross term and $\epsilon_R=10^{-10}$. \label{fig:cmod_zonal}}%
\end{figure}

Same inverse-and-forward test is also performed in a C-Mod-like lower single null configuration with resolution $(n_x,n_y)=(68,128)$ as illustrated in Figures~\ref{fig:cmod_zonal} and~\ref{fig:laplace2d_conv}(b). 
Similar to the shift-circular case, the axisymmetric field solver is able to calculate electrostatic potential $\phi$ and reproduce the right-hand-side of Equation~(\ref{eq:laplace2d}) in a good precision although in this case, error seems to concentrate at the far private flux region (likely due to the choices of initial profiles and boundary conditions).
Interestingly, in the convergence test, no significant discrepancy of the number of iterations with and without the cross term is observed.

\begin{figure}[!ht]
\centering
\includegraphics[width=0.7\linewidth]{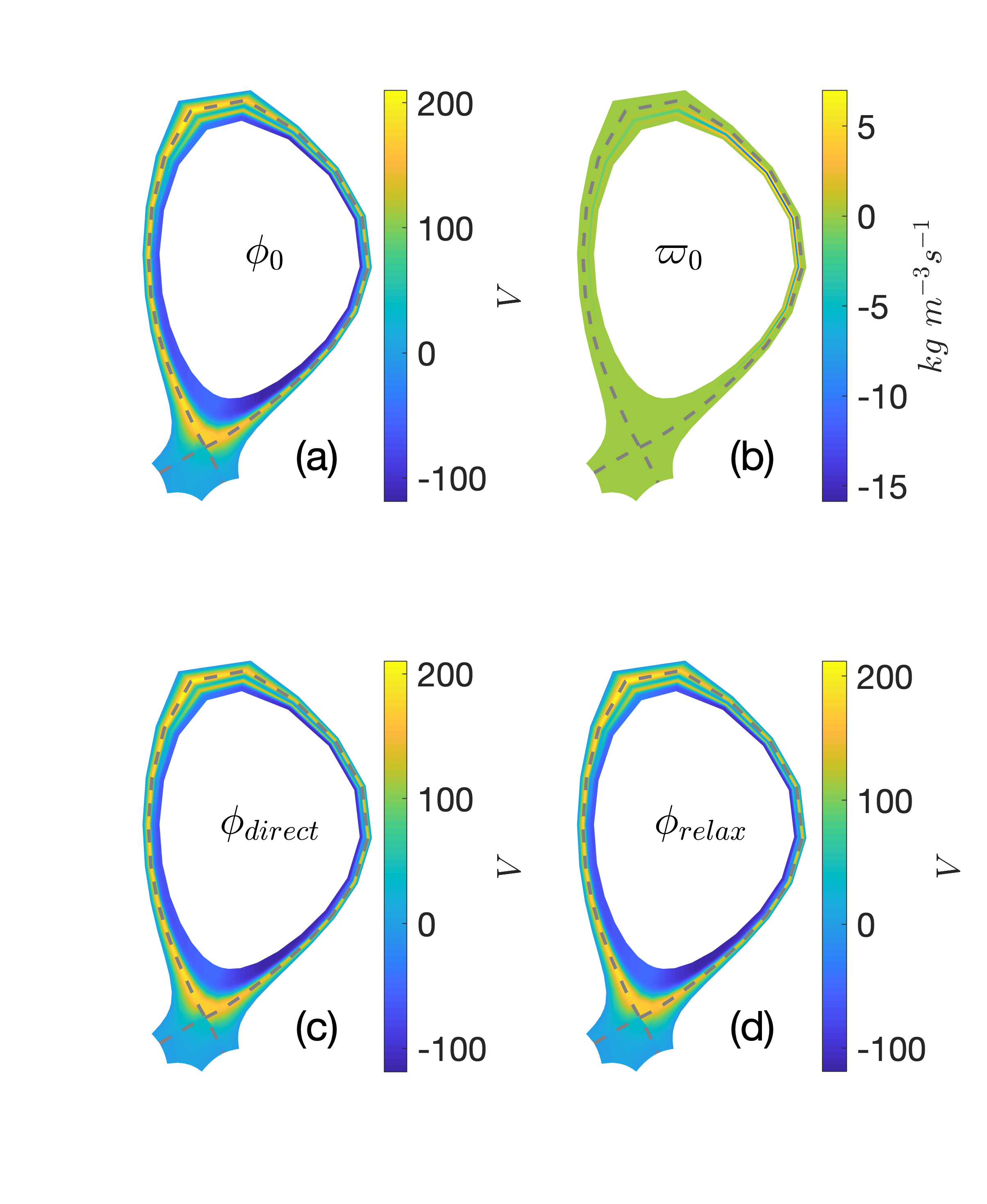}%
\caption{Initial (a) electrostatic potential and corresponding (b) vorticity v.s. the solutions from (c) direct axisymmetric field solver and (d) relaxation method for a lower single null configuration. \label{fig:zonal_bench}}%
\end{figure}

The new axisymmetric field solver is further benchmarked against relaxation method~\cite{li2018calculation} currently implemented in BOUT++'s transport model. As plotted in Figure~\ref{fig:zonal_bench}, for a given initial potential and vorticity profiles $\phi_0,\varpi_0$ , both methods are able to recover the initial potential profile.
The overall small discrepancy between two solutions is likely due to the slightly shifted location of boundary condition implementation -- while relaxation method enforces boundary condition at the boundary grid points, the axisymmetric field solver enforces boundary condition at the boundary cell centers. 

\begin{figure}[!ht]
\centering
\includegraphics[width=\linewidth]{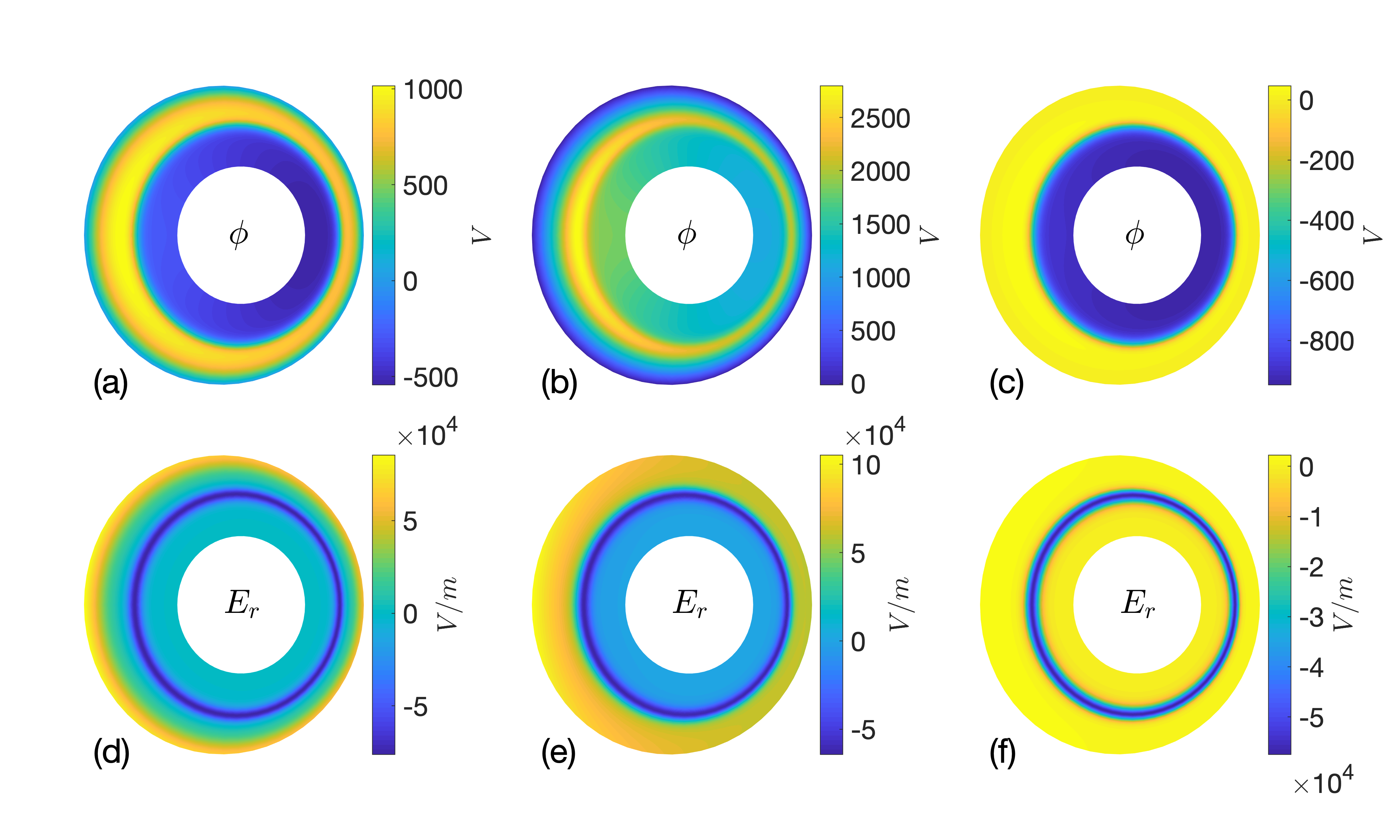}%
\caption{Solutions of axisymmetric electrostatic potentials and the corresponding radial electric fields assuming vorticity $\varpi=0$ for different vorticity expression: (a,d) complete form~(\ref{eq:vort_ff}), (b,e) without cross term~(\ref{eq:vort_bos1}), (c,f) without density dependence~(\ref{eq:vort_bos1}). \label{fig:cbm_bos}}%
\end{figure}

Authors would like to point out that the new field solver is able to invert the a more accurate form of vorticity, e.g., including the $n=0$ component of the cross term contributed from axisymmetric ion density and electrostatic potential in Equation~(\ref{eq:vort_ff}). For years, edge turbulence models either drops the crossing term $\nabla_\perp\phi\cdot\nabla_\perp n_i$ and/or the ion density dependence in front of $\nabla_\perp^2\phi$ term (i.e., so called Boussinesq approximation) in order to simplify the expression of vorticity, hence the numerical calculation,
\begin{align}
    \label{eq:vort_bos1}
    \varpi&\approx\frac{n_i}{B}\left(\nabla_\perp^2\phi+\frac{\alpha_{di}}{n_i}\nabla_\perp^2 P_i\right)\\
    \label{eq:vort_bos2}
    \varpi&\approx\frac{1}{B}\left(\nabla_\perp^2\phi+\alpha_{di}\nabla_\perp^2 P_i\right)
\end{align}
However, the validity of Boussinesq approximation in global edge turbulence simulations is never rigorously justified, especially considering that radial density profile $n$ may have a steep gradient at the pedestal region.
Recent studies suggest that Boussinesq approximation has limited influence ($\sim 10\%$) on the overall turbulent fluctuation level with~\cite{zhu2017global2} and without~\cite{stegmeir2019global} hot ion in a circular geometry. Our preliminary tests indicate that the cross term and the $1/n_i$ dependence do influence the inverted axisymmetric electrostatic potential and the corresponding radial electric field (i.e., zonal field) as shown in Figure~\ref{fig:cbm_bos}. A through study of the impact of Boussinesq approximation on edge instability, turbulence, flow generation and saturation is subject of future research.

\section{Flux driven source}\label{sec:source}
In the edge turbulence simulations, plasma is constantly transported outward once instability is excited, it is therefore reasonable to refuel particles and energy from the core side in order to prevent the collapse of density and temperature profiles in the transport time scale simulations.
Meanwhile, plasma sources are also necessary to simulate certain transient events, e.g., the thermal quench phase of tokamak disruption when the hot core plasma is pushed towards edge in a very short time-scale (~100 $\mu s$).

The newly added sourcing options therefore largely expands the range of physics problems that BOUT++ edge model is able to investigate. A good demonstration of this new capability is the preliminary study of divertor heat load response to excessive heating which is an important issue in tokamak disruption mitigation. 

\begin{figure}[!ht]
\centering
\includegraphics[width=0.5\textwidth]{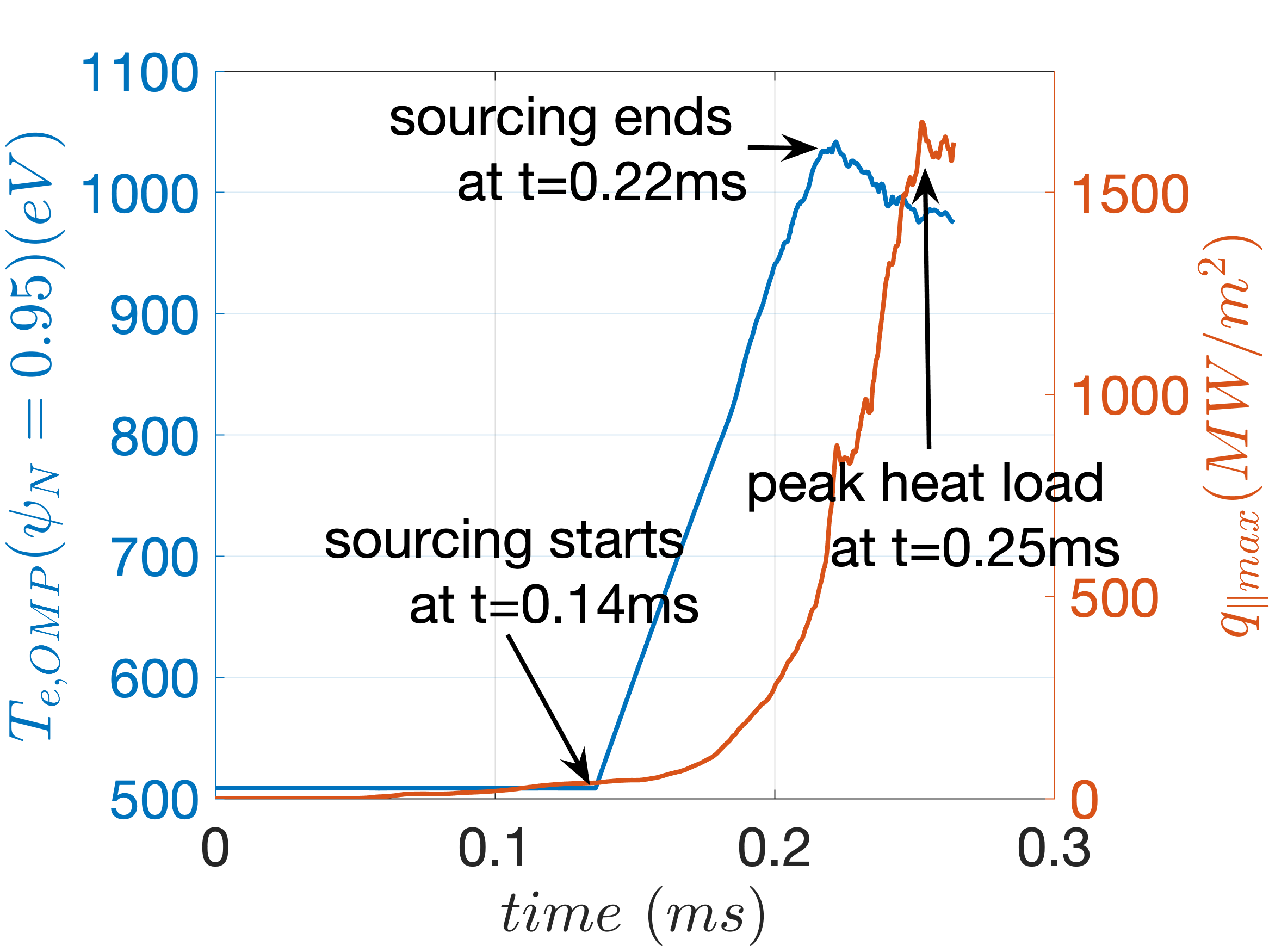}
\caption{Outboard midplane electron temperature at $\psi_N=0.95$ and maximum outer divertor heat load evolution.}
\label{fig:d3d_source1}
\end{figure}

\begin{figure}[!ht]
\centering
\includegraphics[width=\textwidth]{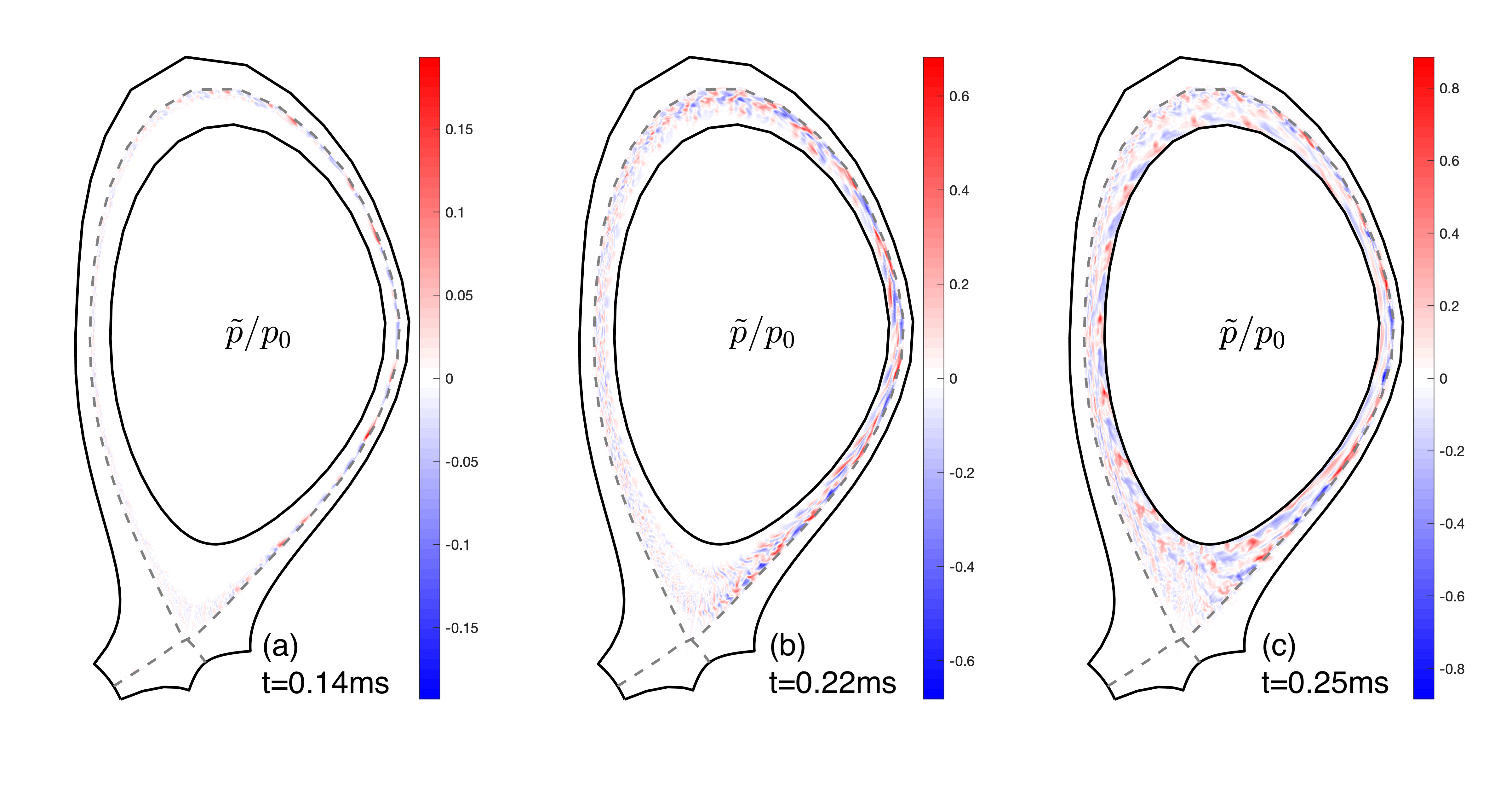}
\caption{Poloidal snapshots of normalized plasma pressure perturbation at three different time, (a) $t=0.14ms$, i.e., the beginning of energy injection, (b) $t=0.22ms$, i.e., the end of energy injection and (c) $t=0.25ms$ which is roughly at when the peak heat load is reached.}
\label{fig:d3d_source2}
\end{figure}

A nonlinear electromagnetic simulation is first carried out in a lower single null DIII-D H-mode plasma without sourcing for the first 0.14 $ms$. For the next 80 $\mu s$, 1 $GW$ power, or 80 $kJ$ energy in total (roughly 15$\%$ of the total stored plasma thermal energy) is injected to the pedestal top region ($0.85<\psi_N<0.96$) to mimic intensive energy flux coming from the core in the thermal quench phase. For simplicity, heating power is equally partitioned between electrons and ions, and assumed to deposit uniformly in the poloidal and toroidal directions while take a Gaussian shape in the radial direction. At $t=0.22~ms$, energy source is turned off. Flux-limited parallel heat flux model is used in this particular simulation.
Figure~\ref{fig:d3d_source1} plots the time history of outboard midplane electron temperature $T_{e,OMP}$ at the pedestal top ($\psi_N=0.95$) and the corresponding peak magnitude of the heat flux at outer divertor target plate.
A monotonic, almost linear increasing of $T_{e,OMP}$ is observed once the power injection starts, followed by a much slower decreasing phase once the source is turned off. The divertor heat flux, on the other hand,  takes a while to response to the change of pedestal temperature. The 30 $\mu s$ lag (e.g., between the peak temperature and peak divertor heat load) is mainly due to the parallel conduction process where the plasma inside the separatrix first enters the SOL primarily near the outboard midplane and then is transported to the divertor target.
In this run, doubled pedestal top temperature yields more than 40 times larger heat flux and doubled heat flux width. The dramatic increasing of heat load, as well as the broadening of heat flux width, are caused by the enhanced turbulence activity at the tokamak edge as illustrated in Figure~\ref{fig:d3d_source2}.
Before the intensive energy injection, the system is marginally stable to the intermediate-$n$ peeling-ballooning modes localized at the steep gradient region right inside the separatrix (Figure~\ref{fig:d3d_source2}(a)). Once the heating is on, elevated temperature profiles steepen edge plasma pressure gradients and destabilize these previously marginal stable intermediate-$n$ modes. Despite these modes are strongly sheared by the edge $E_r$ field, they are still able to spread both inward to the pedestal top and outward across the separatrix into the scrape-off-layer, and therefore, amplifies cross-field heat transport(Figure~\ref{fig:d3d_source2}(b)). As the pedestal plasma is substantially heated, low-$n$ modes are eventually excited near the pedestal top, providing a more violent yet effective turbulent transport channel and resulting an even larger heat load on the divertor target plate (Figure~\ref{fig:d3d_source2}(c)).
Although the details of the underlying physics in process is yet to be explored, the versatile flux-driven source option provides a unique opportunity to study edge plasma dynamics in this pulsed over-driven setting. 

\section{Summary} \label{sec:sum}
The drift-reduced two fluid six-field turbulence model in BOUT++ framework is perhaps the most widely used tokamak edge turbulence code in the magnetic fusion community. It evolves plasma density, electron and ion temperature, ion parallel velocity, electrostatic potential and parallel magnetic flux in a realistic tokamak magnetic configuration. In this paper, we present in detail the current status of this model in a self-contained manner, including the model equations, normalization, coordinate system, operators, boundary conditions and the recently added features such as flux-driven source, Landau fluid closure on parallel heat flux and the axisymmetric field solver. These new features greatly extend the model capability to simulate transport time scale plasma dynamics (i.e., shear flow generation, turbulence saturation and profile evolution) and to study critical edge problems such as divertor heat flux solutions. 

It should be noted that BOUT++ remains an active project. Numerical and physics models within BOUT++ are continuously refined. For example, in the numerical side, GPU-capability and a new 3D field solver are under development; while on the physics side, a comprehensive fluid neutral model and a ML-enabled surrogate model~\cite{ma2020machine} of an even more sophisticated kinetic Landau fluid closure~\cite{wang2019landau} are also on the way.

\section*{Acknowledgments}
We thank Drs. J. Chen, A. Dimits, B. Dudson, N. Li, C. Ma, J. Omotani and M. Umansky for helpful discussions.
This work is performed under the auspices of the U.S. Deportment of Energy (US DOE) by Lawrence Livermore National Laboratory under Contract DE-AC52-07NA27344 and supported by the US DOE Tokamak Disruption Simulation (TDS) Scientific Discovery Through Advanced Computing (SciDAC) program. This research used resources of the National Energy Research Scientific Computing Center, a DOE Office of Science User Facility supported by the Office of Science of the US DOE under Contract No. DE-AC02-05CH11231. LLNL-JRNL-812150.





\bibliographystyle{elsarticle-num}
\bibliography{6f_landau}







\end{document}